\documentclass[12pt,a4paper]{article}

\usepackage[T1]{fontenc}
\usepackage[hmarginratio=1:1,top=32mm,columnsep=20pt]{geometry}
\usepackage{mathdots}
\usepackage{upgreek}
\usepackage{color}
\definecolor{dark-gray}{gray}{0.20}
\definecolor{gray}{gray}{0.30}
\definecolor{light-gray}{gray}{0.80}
\definecolor{dark-red}{rgb}{0.7,0,0}
\definecolor{dark-green}{rgb}{0.1,0.4,0}
\definecolor{dark-blue}{rgb}{0.3,0.3,0.7}
\definecolor{light-blue}{rgb}{0.8,0.8,1}
\definecolor{purple}{rgb}{0.5, 0, 0.5}
\DeclareFontFamily{OMS}{rsfs}{\skewchar\font'60}
\DeclareFontShape{OMS}{rsfs}{m}{n}{<-5>rsfs5 <5-7>rsfs7 <7->rsfs10 }{}
\DeclareSymbolFont{rsfs}{OMS}{rsfs}{m}{n}
\DeclareSymbolFontAlphabet{\Scr}{rsfs}

\usepackage{cite}
\usepackage{hyperref}
\hypersetup{
	colorlinks=true,
	linkcolor=dark-blue,
	citecolor=dark-red,
	urlcolor=dark-green,
	linktoc=page
}

  \usepackage{a4wide}
  \usepackage{latexsym}
  \usepackage{epsf}
  \usepackage{bbm}
  \usepackage{graphicx}
  \usepackage{amsmath,amssymb,amsthm}
  \usepackage{verbatim}

\newcommand{\bbm}{\left(\begin{matrix}}
\newcommand{\ebm}{\end{matrix}\right)}
\newcommand{\bea}{\begin{eqnarray}}
\newcommand{\eea}{\end{eqnarray}}
\newcommand{\be}{\begin{equation}}
\newcommand{\ee}{\end{equation}}

\renewcommand{\d}{\textrm{d}}

\newcommand{\SL}{\mathop{\rm SL}}
\newcommand{\GL}{\mathop{\rm GL}}
\newcommand{\SO}{\mathop{\rm SO}}

\newcommand{\SU}{\mathop{\rm SU}}

\begin{document}

\numberwithin{equation}{section}

\begin{center}

{\LARGE {\bf  The holographic dual to \\ \vspace{0.5 cm} supergravity instantons in $\rm AdS_5\times S^5/\mathbb{Z}_k$ }}  \\

\vspace{1.5 cm} {\large  S.~Katmadas$^a$, D.~Ruggeri $^{b,c}$, M.~Trigiante$^{b,c}$ and  T.~Van Riet$^a$ }\footnote{{ \upshape\ttfamily daniele.rug@gmail.com, mario.trigiante@polito.it,  stefanos.katmadas, thomas.vanriet @kuleuven.be
 } }\\
\vspace{0.65 cm}
${}^a$Instituut voor Theoretische Fysica, K.U. Leuven,\\
Celestijnenlaan 200D B-3001 Leuven, Belgium

{ ${}^b$Department of Applied Science and Technology, Politecnico di Torino, \\ C.so Duca degli Abruzzi, 24, I-10129 Torino, Italy\\ \vspace{.15 cm}
}
{ ${}^c$ INFN, Sezione di Torino,
via P. Giuria 1, 10125 Torino, Italy\\ \vspace{.15 cm}
}

\vspace{2cm}

{\bf Abstract}
\end{center}

{\small We investigate the holographic dual to supergravity instanton solutions in $\rm AdS_5\times S^5/\mathbb{Z}_k$, which  are described entirely in terms of geodesics on the AdS moduli space. These instantons are expected to be holographically dual to instantons in an $\mathcal{N}=2$ necklace quiver gauge theory in four dimensions with $k$ gauge nodes, at large N. For the supersymmetric instantons we find a precise match between the moduli-dependent parts of the on-shell actions and the vevs of $\text{Tr}[F^2]$ and $\text{Tr}[F\wedge F]$ computed on both sides of the duality. This correspondence requires an exact identification between the massless supergravity scalars and the dual gauge couplings which we give in detail. We also find a candidate for the supergravity dual of a quasi-instanton in the quiver theory, which is a non-supersymmetric extremal solution.

\thispagestyle{empty}
\setcounter{page}{0}
\setcounter{tocdepth}{2}
\newpage
\tableofcontents

\section{Introduction}
In this paper we investigate the holographic dual description of supergravity instantons in the $\rm AdS_5\times S^5/\mathbb{Z}_k$ vacuum of IIB supergravity. Such instanton solutions were recently constructed in \cite{Ruggeri:2017grz} building on earlier work in \cite{Hertog:2017owm}. As explained in those references the instanton solutions can be found from 5-dimensional half-maximal gauged supergravity truncated to the moduli space of the AdS vacuum. This moduli space has been studied in quite some detail in \cite{Corrado:2002wx, Louis:2015dca} and shown to be the coset:
\be\label{coset2}
\mathcal{M}_{\text{moduli}} = \frac{\SU(1,k)}{{\rm S[U(1)\times U(k)]}}\,.
\ee
Since we wish to describe instantons, we need to know the moduli space of the AdS vacuum in Euclidean signature. The Wick rotation of space-time induces a Wick rotation of the moduli space which is neither unique nor fixed by Euclidean supersymmetry.  But given the higher-dimensional and holographic interpretation of the scalars it can be fixed uniquely as follows. Since IIB string theory on $\rm AdS_5\times S^5/\mathbb{Z}_k$ is expected to be the holographic dual of the $\mathcal{N}=2$ necklace quiver gauge theories with $k$ nodes \cite{Kachru:1998ys}, the moduli space of the AdS vacuum is then supposed to be dual to the space of exactly marginal couplings, a.~k.~a.~the conformal manifold. The marginal couplings in the necklace quiver gauge theories are $k$ complexified couplings, of which the real parts correspond to $\theta$-angles and hence get Wick-rotated with an $i$-factor. The scalars that are dual to these $\theta$ angles should be scalars that enjoy a (classical) shift symmetry (axions). The manifold (\ref{coset2}) has exactly $k$ Abelian isometries that act as shifts of $k$ real scalars. This fixes the Wick-rotation uniquely to \cite{Hertog:2017owm}
\be\label{coset1}
\mathcal{M}_{\text{moduli}} = \frac{\SL(k+1,\mathbb{R})}{\GL(k,\mathbb{R})}\,.
\ee

If one restricts to instanton solutions with spherical symmetry (i.e.~respecting the rotational $SO(5)$ symmetry of Euclidean $\rm AdS_5$) for simplicity, then the scalar field equations of motion reduce to geodesic equations on $\mathcal{M}_{\text{moduli}}$ and the Einstein equations of motion decouple from the scalar fields into a universal form, \cite{Breitenlohner:1987dg, Bergshoeff:2008be, Hertog:2017owm}. The metric is then only sensitive to the constant velocity of the geodesics (in an affine parametrization), which we denote as $c$
\be\label{velocity}
G_{\mathcal{IJ}}\dot{\phi}^{\mathcal{I}}\dot{\phi}^{\mathcal{J}} = c\,,
\ee
where $\phi^{\mathcal{I}}$ denote the moduli, $G_{\mathcal{IJ}}$ is the metric on (\ref{coset1}) and a dot is a derivative with respect to the affine coordinate. Not surprisingly the description of geodesics on a coset space like (\ref{coset1}) can be understood entirely using group theory \cite{Ruggeri:2017grz}. The properties and meaning of geodesics depend strongly on whether they are time-like ($c>0$), space-like ($c<0$) or null ($c=0$).

When $c<0$ the instantons are called \emph{super-}extremal and the geometry describes a smooth two-sided wormhole that asymptotes to AdS on both sides \cite{Gutperle:2002km}. A very explicit and concrete embedding of such wormholes was found inside $\rm AdS_5\times S^5/\mathbb{Z}_k$ when $k>1$ \cite{Hertog:2017owm}. When $k=1$ the scalars are singular and the corresponding wormholes are not considered as physical \cite{Bergshoeff:2005zf}. These Euclidean ``axionic'' wormholes have a long history in cosmology, QCD, holography and quantum gravity (see for instance \cite{Giddings:1987cg, Coleman:1988cy, Lavrelashvili:1987jg, Bergshoeff:2004pg,  ArkaniHamed:2007js, Montero:2015ofa, Hebecker:2016dsw, Hertog:2017owm, Alonso:2017avz} ) which was reviewed in the comprehensive paper \cite{Hebecker:2018ofv}.

When $c>0$ the instantons are called \emph{sub-}extremal and the geometry corresponds to a singular ``spiky'' deformation of AdS. The holographic description of these instantons is unclear but for $c$ small enough a suggestion was made in \cite{Bergshoeff:2005zf} in the case of $\rm AdS_5\times S^5$. In this paper we find that the same interpretation holds water as well in $\rm AdS_5\times S^5/\mathbb{Z}_k$.

The focus of our paper is on the extremal instantons, for $c=0$.  When $k>1$ it was found that these instantons come into two families, supersymmetric or not, whereas for $k=1$ there are only supersymmetric solutions. One can expect that the supersymmetric instantons are the easiest to understand. When $k=1$ these instantons are simply the D(-1)/D3 bound states zoomed in near the horizon of the D3. Their holographic map to supersymmetric instantons in $\mathcal{N}=4$ Yang-Mills theory has been established in the early days of AdS/CFT \cite{Balasubramanian:1998de, Chu:1998in, Bianchi:1998nk,Kogan:1998re, Banks:1998nr, Dorey:1998xe, Dorey:1998qh, Dorey:1999pd, Green:2002vf, Belitsky:2000ws}. For $k>1$ much less is known. A preliminary discussion can be found in \cite{Hollowood:1999bm} and an in-depth study of the instanton moduli space (and other properties) was performed in \cite{Argurio:2007vqa, Argurio:2012iw}. It is the aim of this paper to present the detailed map between the on-shell actions on both sides of the correspondence and the dual one-point functions for the marginal operators. As it turns out, this map works nicely, apart from a moduli-independent constant, for the supersymmetric instantons and leads to a clear description of the holographic dictionary between the massless scalars and the dual marginal operators, which to our knowledge has not been constructed before.\footnote{We leave the understanding of the constant difference between the two actions to a future investigation.} For the non-supersymmetric instantons we present evidence, using the one-point functions, that they correspond to the so-called ``quasi-instantons'' \cite{Imaanpur:2008jd} of quiver gauge theories,  although the on-shell actions do not map to each other. We speculate on why that happens.

The remainder of this paper is organised as follows. In section \ref{sec:general} we review the results of \cite{Ruggeri:2017grz} on instantons in $\rm AdS_5$. We then proceed in section \ref{sec:instantons} to similarly review some basic properties of instantons in field theory relevant to the holographic map, focussing on the one-point functions, Pontryagin indices and on-shell actions. In section \ref{sec:hol-map} we combine these two pictures to obtain the holographic dictionary between the massless scalars and the dual marginal operators, by analysing in detail the supersymmetric case. We also briefly discuss aspects of the holographic interpretation of the non-supersymmetric extremal and non-extremal instantons in $\rm AdS_5$. We conclude in section \ref{sec:discussion}, with some comments on the holographic map in more general settings. In Appendices \ref{CHk} and \ref{Solvpar} we give more details on the geometry of the moduli space. Finally in Appendix \ref{App:IIB} we compare the geometries of the moduli spaces of two different theories: Type IIB on $AdS_5\times S^5/\mathbb{Z}_2$ and on $AdS_5\times T^{1,1}$. In particular we show that the corresponding metrics differ by terms which are of higher order in the values of the Type IIB 2-forms on the internal 2-cycles.

\section{Reviewing supergravity instanton solutions} \label{sec:general}
\subsection{General comments}
In order to describe deformations of $\rm AdS_5\times S^5/\mathbb{Z}_k$ for which the dual sources that are possibly turned on are exactly marginal, one should restrict to the moduli space of the vacuum. Instantons satisfy this requirement, since they should correspond to deformations that can be interpreted entirely as non-trivial vacuum configurations of the undeformed theory with non-zero vev's for the operator that corresponds to the Lagrangian itself, so $\text{Tr}[F^2]+\ldots$ and $\text{Tr}[F\wedge F]$, which are exactly marginal. Hence a good starting point is to consider the gauged supergravity obtained by truncating the KK modes of Type IIB supergravity on $S^5/\mathbb{Z}_k$ \cite{Corrado:2002wx}, restricted to the moduli space of its supersymmetric vacuum. If then only scalars are switched on, we end up with an effective action of the following form:\footnote{Here we point out that, in the background under consideration, the relation between $\kappa_5^2$ and $\kappa_{10}^2$ is the following:
\begin{equation}
\frac{1}{\kappa_5^2}=\frac{\ell^5\,{\rm Vol}(S^5/\mathbb{Z}_k)}{\kappa_{10}^2}=\frac{\ell^5\,{\rm Vol}(S^5)}{k\,\kappa_{10}^2}=\frac{\ell^5\,\pi^3}{k\,\kappa_{10}^2}\,,
\end{equation}
where we have used the property: ${\rm Vol}(S^5/\mathbb{Z}_k)={\rm Vol}(S^5)/k=\pi^3/k$.}
\be\label{5Daction}
S = -\frac{1}{2\kappa_5^2} \int \sqrt{|g_5|}\Bigl(R_5 - \tfrac{1}{2}G_{\mathcal{IJ}}\partial\phi^{\mathcal{I}}\partial\phi^{\mathcal{J}}  - \Lambda \Bigr)\,,
\ee
with the $\phi^{\mathcal{I}}$ coordinates on the moduli space $\mathcal{M}_{\text{moduli}}$, $G_{\mathcal{IJ}}$ its canonical metric and $\Lambda$ the negative minimum of the scalar potential in the supersymmetric AdS vacuum.   If we restrict to instanton solutions with spherical symmetry, the metric Ansatz is given by
\begin{equation}
\d s_5^2 = f(r)^2\d r^2 + a(r)^2\d\Omega_4^2\,,
\end{equation}
and the moduli only depend on Euclidean time $r$.  In the gauge $f=a^4$, $r$ is an affine parametrization of the geodesic curves that are being traced out by the scalars, such that $\dot{\phi}^{\mathcal{I}}$ in equation (\ref{velocity}) means $\dot{\phi}^{\mathcal{I}}=\d\phi^{\mathcal{I}}/\d r$. The Einstein equations are then equivalent to the following Hamiltonian constraint
\begin{equation}\label{firstorder}
\frac{\dot{a}^2}{f^2} = \frac{c}{24}a^{-6} + \frac{a^2}{\ell^2} + 1\,,
\end{equation}
where $\ell$ is the AdS length scale, defined as $\Lambda=-12/\ell^2$, and $c$ is the integration constant appearing in \eqref{velocity}. We therefore see that the metric is determined by $c$ only.  When $c=0$ the metric is that of pure Euclidean AdS. This is due to the vanishing of the total energy momentum of the scalar fields, which is possible because of the indefinite sigma model metric.  The extremal instantons can straightforwardly be extended to non-spherical solutions by replacing $r$ by a general harmonic function that is not radially symmetric. We shall use the symbol $\tau$ to denote the harmonic function.\par When the variable $\tau$ is used for extremal solutions $c=0$ it can mean any harmonic but for non-extremal solutions $c\neq 0$ we have the spherically symmetric harmonic on Euclidean $\rm AdS_5$ in mind:
 \be
 \partial_r\left(f^{-1}a^4\partial_r \tau(r)\right) =0\,.
 \ee
The most general harmonic function $H$ with a single center is given by
\be
H(z, \vec{x}) = \alpha F^{-3}\left( \left(1 - \frac{2F^2}{z_0^2}\right)\sqrt{1+\frac{F^2}{z_0^2}}\right) + \beta\,,\label{Hharm}
\ee
with $\alpha, \beta$ constants\footnote{We fix $\alpha$ and $\beta$ such that for the spherically symmetric harmonic we simply have $H=r$ in the gauge $f=a^4$} and $F$
is the $\SO(1,5)$ invariant function:
\be
F(z, \vec{x}) = \frac{\sqrt{[(z_0-z)^2 + |\vec{x}-\vec{x}_0|^2][(z_0+ z)^2 + |\vec{x}-\vec{x}_0|^2]}}{2z}\,,\label{FFunc}
\ee
with $\vec{x}, z$ Poincar\'e coordinates for which EAdS  metric is given by $\d s^2 = \ell^2 z^{-2}\Bigl(\d z^2 + \d\vec{x}^2\Bigr)$. The singularity of $H$ at $z=z_0, \vec{x}=\vec{x}_0$ can be interpreted as the position of the instanton and is free. Therefore the whole of $\rm EAdS_5$ is part of the instanton moduli space. The specific choice $z_0=\ell, \vec{x}_0=0$ can be thought of as the original spherically-symmetric solution, where $H\sim \tau$. The most general extremal solution now consists in taking an arbitrary superposition of harmonics with singularities at different positions. These can be thought of as multi-centered instantons.

 \subsection{Geodesic solutions}
We now turn to the explicit expressions for the moduli-geodesics in the specific model considered in this paper, given in \eqref{coset1}. Following \cite{Hertog:2017owm,Ruggeri:2017grz} we introduce $2k$ real coordinates on $\mathcal{M}_{\text{moduli}}$, denoted by $U , a, \zeta^i ,\,\tilde{\zeta}_i$ where $i=1,\ldots, k-1$ \footnote{With respect \cite{Ruggeri:2017grz}, we make, for notational convenience, the following rescaling
\begin{equation}
\zeta^i \,\rightarrow\,\frac{{\zeta}^i}{\sqrt{2}}\quad,\quad \tilde{\zeta}_i \,\rightarrow\,\frac{\tilde{\zeta}_i}{\sqrt{2}}\quad,\quad L_i \,\rightarrow\,\frac{L_i}{\sqrt{2}}\,.
\end{equation}
}. These coordinates define the so-called \emph{solvable parametrization} of the coset, see Appendix \ref{Solvpar}. The metric on the moduli space can then be written as:
\begin{equation}\label{metric}
\d s^2=4 \d U^2-{e^{-4U}}{\Scr N}^2\,+\,{e^{-2U}}\sum_{i=1}^{k-1}[(\d\zeta^i)^2-(\d\tilde{\zeta}_i)^2]\,,
\end{equation}
where ${\Scr N}$ is the  one-form, $
{\Scr N}\equiv da + (\sum_i \zeta^i \d \tilde{\zeta}_i  - \tilde{\zeta}_i\d\zeta^i)/2 $.
In contrast, the metric on the moduli space of Lorentzian AdS which (somewhat confusingly) has Euclidean signature, can be obtained trivially from the above metric by analytically continuing ${\Scr N}$ and $(d\tilde{\zeta}_i)$ to imaginary values, thus flipping the negative signs in the metric \eqref{metric}.\par Prior to the Wick rotation, the moduli space (\ref{coset2}) is also called complex hyperbolic space and is denoted by $\mathbb{C}H^k$. For the reader's convenience we summarize in Appendix \ref{CHk} the relevant parametrizations of this space and their relations. \par
As opposed to the solvable parametrization of the Riemannian scalar manifolds occurring in Lorentzian supergravities, the solvable parametrization of a pseudo-Riemannian manifold (such as $\mathcal{M}_{{\rm moduli}}$ in (\ref{coset1})) is in general not global but only covers a local patch in the space. Such parametrization however defines the physical fields $\phi^{\mathcal{I}}=\,U ,\, a,\, \zeta^i ,\,\tilde{\zeta}_i$, each of which corresponds to string excitations on the chosen background. The precise string-interpretation of the solvable coordinates will be given later. We therefore refer to the corresponding local patch as the \emph{physical patch}. At its boundary the fields are no longer well defined and, in particular, $e^{-2 U}\rightarrow 0$. A solution described by a given geodesic can be singular although its space-time geometry is regular, as it is the case for the space-like geodesics describing wormholes. This occurs if the arc of the geodesic described by the solution crosses the border of the physical patch, where some of the moduli diverge. This never happens for light-like and time-like geodesics as the affine parameter $\tau$ runs from $0$ to $\infty$, but only for the space-like ones. Therefore the regularity condition for wormholes amounts to requiring that the segment of the corresponding geodesic on $\mathcal{M}_{{\rm moduli}}$, described when going from side to side of the solution, never crosses the boundary of the physical patch.

It is also important to point out that the solvable parametrization $U ,\, a,\, \zeta^i ,\,\tilde{\zeta}_i$ is manifestly covariant only with respect to the  ${\rm GL}(k-1)$ subgroup of the ${\rm GL}(k)$ isotropy group, which acts linearly on the $\zeta^i ,\,\tilde{\zeta}_i$ coordinates leaving $a$ and $U$ invariant. The correspondence with string excitations, however, will require considering  the combinations
\begin{equation}
 \delta\equiv e^{2U}+\frac{1}{2}\sum_{i=1}^{k-1} (\zeta^i)^2\,, \qquad  \tilde{a}\equiv a - \frac{1}{2}\sum_i\tilde{\zeta}_i\zeta^i\,,
\end{equation}
to be identified as
\begin{equation}
 \delta \sim e^{-\phi}=1/g_s\,, \qquad \chi \sim - \tilde{a}\,,
\end{equation}
where $\phi$ and $\chi$ being the ten-dimensional type IIB dilaton and axion fields, respectively. These combinations break ${\rm GL}(k-1)$ down to ${\rm O}(k-1)$, so that the string interpretation of our solutions is made in a ${\rm O}(k-1)$-covariant frame. The symmetry group $\mathbb{Z}_k$ of the $A_{k-1}$-quiver is contained in ${\rm O}(k-1)$ as we shall show in Sect. \ref{sec:hol-map}. We refer the reader to Appendix \ref{Solvpar} for more mathematical details about the solvable parametrization and its manifest symmetries.

The geodesic solutions can most easily be constructed using the exponential map $M = M(0) \exp(2Q \tau)$, with $Q$ an element of the Lie algebra of the coset, $\tau$ the affine coordinate and $M$ a matrix, built from the coset representative $\mathbb{L}$ in the solvable gauge.  The details can be found in \cite{Ruggeri:2017grz}. Here we simply present the solutions in terms of the scalars. The solutions are all characterized by $2k$ ``charges'' denoted $p_{\alpha}, m_{\beta}$ with $\alpha, \beta=0\ldots k-1$ that obey
\be \label{mu}
c = 4\,\vec{m}\cdot \vec{p}\,.
\ee
For simplicity we explicitly write the geodesics through the origin, $O$, of the moduli space. The general solutions that do not pass through $O$  are obtained by acting on the ones originating in $O$ by means of shift-like isometry transformations:
\begin{align}
 U & \rightarrow U + U(0)\,,\nonumber\\
\tilde{\zeta} & \rightarrow \tilde{\zeta}e^{U(0)} + \tilde{\zeta}(0)\,,\qquad
\zeta  \rightarrow \zeta e^{U(0)} + \zeta(0)\,,\nonumber\\
a & \rightarrow a e^{2U(0)} + \frac{\zeta\tilde{\zeta}(0)e^{U(0)}}{2} - \frac{\tilde{\zeta}\zeta(0)e^{U(0)}}{2} + a(0)\,.\label{shifstsym}
\end{align}
The above transformations are isometries in $\SL(k+1,\mathbb{R})$ that act transitively on the coset. In fact they are generated by the solvable Lie algebra $Solv$ parametrized by the scalar fields in the solvable parametrization. We shall generically describe the geodesic passing through a point $\phi_0=(\phi_0^{\mathcal{I}})=(U(0),\,\zeta^i(0),\,\tilde{\zeta}_i(0),\,a(0))$ of the moduli space at $\tau=0$ using the compact notation:
\begin{equation}
\phi^{\mathcal{I}}=\phi^{\mathcal{I}}(\tau,\,\phi_0)\,\,\,,\,\,\,\,\,\,\,\,\,
\phi^{\mathcal{I}}(\tau=0,\,\phi_0)=\phi_0^{\mathcal{I}}\,.\label{gengeo}
\end{equation}

Note that the boundary of Euclidean AdS is at $\tau=0$. So the geodesics originate at the boundary (UV) and flow to the interior as $\tau\rightarrow \infty$.

\subsubsection*{Lightlike geodesics}
When $c=0$ we have:
\begin{align}
U\,&=\,\frac{1}{2}\,\log\left[\frac{1}{\left(1-\tau p_0\right)\left(1-\tau m_0\right)}\right]\quad,\nonumber\\
\zeta^i\,&=\,-\tau\,\left(\,\frac{p_i}{\left(1- \tau p_0\right)}\,+\,\frac{m_i}{\left(1-\tau m_0\right)}\,\right)\quad,\nonumber\\
\tilde{\zeta}_i\,&=\,-\tau\,\left(\,\frac{p_i}{\left(1-\tau p_0\right)}\,-\,\frac{m_i}{\left(1-\tau m_0\right)}\,\right)\quad,\nonumber\\
a\,&=\,-\frac{1}{\left(1-\tau p_0\right)}\,+\,\frac{1}{\left(1-\tau m_0\right)}\quad\label{gensol4},
\end{align}
where $i=1,\dots, k-1$.  The condition $c=0$ implies that the Noether charge matrix $Q$ is nilpotent  either of  degree 2 ($Q^2=0$) or degree 3 ($Q^3=0$) \cite{Ruggeri:2017grz}. The degree 2 solutions are supersymmetric and imply that either all $p$'s vanish or all $m$'s vanish. Regularity of the solution requires that $p_0 \leq 0$ and $m_0 \leq 0$.

\subsubsection*{Time-like geodesics}
The solutions through the origin with $c\,=\,4\,\vec{p}\cdot \vec{m}\,=\,4\,\mu^2>0$ read:
\begin{align}
U\,&=\,\frac{1}{2}\,\log\left[\frac{\mu^2}{\left(m_0 \sinh\left(\mu \tau\right)-\mu\cosh\left(\mu \tau\right)\right)\left(p_0 \sinh\left(\mu \tau\right)-\mu\cosh\left(\mu \tau\right)\right)}\right]\quad,\nonumber\\
\zeta^i\,&=\,\,\frac{m_i}{m_0-\mu\coth\left(\mu \tau\right)}\,+\,\frac{p_i}{p_0-\mu\coth\left(\mu \tau\right)}\quad,\nonumber\\
\tilde{\zeta}_i\,&=\,-\frac{m_i}{m_0-\mu\coth\left(\mu \tau\right)}\,+\,\frac{p_i}{p_0-\mu\coth\left(\mu \tau\right)}\quad,\nonumber\\
a\,&=\,-\frac{m_0}{m_0-\mu\coth\left(\mu \tau\right)}\,+\,\frac{p_0}{p_0-\mu\coth\left(\mu \tau\right)}\quad\label{gensolnex4},
\end{align}
where $\mu=\sqrt{\vec{m}\cdot \vec{p}}$. Regularity of the solution requires that $p_0 \leq 0$ and $m_0 \leq 0$.

\subsubsection*{Space-like geodesics}\label{INBPS}
If $c\,=\,4\,\vec{p}\cdot \vec{m}\,=\,-4\,\mu^2<0$, we simply replace $\mu\rightarrow i\,\mu$ in the time-like solutions to obtain:
\begin{align}
U\,&=\,\frac{1}{2}\,\log\left[\frac{\mu^2}{\left(m_0 \sin\left(\mu \tau\right)-\mu\cos\left(\mu \tau\right)\right)\left(p_0 \sin\left(\mu \tau\right)-\mu\cos\left(\mu \tau\right)\right)}\right]\quad,\nonumber\\
\zeta^i\,&=\,\frac{m_i}{m_0-\mu\,{\rm cotg}\left(\mu \tau\right)}\,+\,\frac{p_i}{p_0-\mu\,{\rm cotg}\left(\mu \tau\right)}\quad,\nonumber\\
\tilde{\zeta}_i\,&=\,-\frac{m_i}{m_0-\mu\,{\rm cotg}\left(\mu \tau\right)}\,+\,\frac{p_i}{p_0-\mu\,{\rm cotg}\left(\mu \tau\right)}\quad,\nonumber\\
a\,&=\,-\frac{m_0}{m_0-\mu\,{\rm cotg}\left(\mu \tau\right)}\,+\,\frac{p_0}{p_0-\mu\,{\rm cotg}\left(\mu \tau \right)}\quad\label{gensolnex5},
\end{align}
where now $\mu=\sqrt{-\vec{m}\cdot \vec{p}}$.\par
As discussed in \cite{Hertog:2017owm}, the generating solution of regular wormholes corresponds to a space-like geodesic unfolding within the ${\rm SL(2,\mathbb{R})/SO(1,1)}$ submanifold of $\mathcal{M}_{{\rm moduli}}$ spanned by the scalars $U$ and $\tilde{\zeta}_1$. This solution is obtained from the general one by choosing, as the only non-vanishing parameters, $m_0=p_0={\rm f}_0\le 0$ and $m_1=-p_1={\rm f}_1$, with ${\rm f}_1^2>{\rm f}_0^2$, and reads:
\begin{equation}\label{worm-example}
e^{-2U}=\frac{1}{\mu^2}\,(|{\rm f}_0|\,\sin(\mu\tau)+ \mu\,\cos(\mu\tau))^2\,\,,\,\,\,\tilde{\zeta}_1=\frac{2\,{\rm f}_1}{|{\rm f}_0|\,\sin(\mu\tau)+ \mu\,\cos(\mu\tau)}\,,
\end{equation}
where $\mu^2={\rm f}_1^2-{\rm f}_0^2>0$ and $c\,=\,4\,({\rm f}_0^2-{\rm f}_1^2)<0$. The parameter ${\rm f}_0$ represents the component of the velocity vector along the positive signature direction in the pseudo-Riemannian moduli space. The singular points in the moduli space are those in which the geodesic intersects the boundary of the physical patch. At these points $e^{-2U}\rightarrow 0$. We are interested in the geodesic of maximal length $\ell_{{\rm max}}$ between two such points. For fixed ${\rm f}_1$, the maximal length space-like geodesic are obtained by setting ${\rm f}_0=0$.
In this limit $e^{-2U}= \cos(\mu\tau)^2$ and the singular points occur when $\tau=\pm \pi/(2\mu)$, where now $\mu=|{\rm f}_1|$.
The length along such geodesic between the two singular points is readily computed to be
\begin{equation}
\ell_{{\rm max}}=\int_{-\frac{\pi}{2\mu}}^{\frac{\pi}{2\mu}} \sqrt{|c|}\,d\tau\,=\, 2\pi\,,
\end{equation}
where $\sqrt{|c|}=2\mu$.
This value is larger than the actual length $\ell_{{\rm wh}}$ of the arc of the geodesic from side to side of the wormhole, which was computed in \cite{ArkaniHamed:2007js}. The scalar fields are therefore always regular along the solution.
In the limit $|{\rm f}_0|\rightarrow |{\rm f}_1|$, $c\rightarrow 0$ and the instanton becomes extremal non-BPS. The solution discussed in this subsection, and its extremal limit, will be referred to as \emph{intrinsically non-BPS} since there is no choice of the parameters for which it is supersymmetric. Indeed the particular ${\rm SL(2,\mathbb{R})/SO(1,1)}$ submanifold of $\mathcal{M}_{{\rm moduli}}$, spanned by the scalars $U$ and $\tilde{\zeta}_1$, in which the corresponding geodesic unfolds, contains only non-supersymmetric solutions. We shall reconsider this intrinsically non-BPS instanton  later in light of the holographic correspondence.

\subsection{Axion Charge Quantization}\label{acq}
As emphasized earlier, the AdS moduli space contains $k$ axion fields, and the shifts of these fields define $k$ commuting Killing vectors. The scalar fields that posses these shift symmetries are the $\tilde{\zeta}_i$ and $\tilde{a}$. The latter scalar combination is defined as
\begin{equation}
\tilde{a}  =  a - \frac{1}{2}\sum_i\tilde{\zeta}_i\zeta^i\,.
\end{equation}
The charges under these symmetries are called axion charges and should be quantised. This in turn implies that certain combinations of $m$' and $p$'s are quantised.
 It is useful to adopt the following collective notation for describing the axion fields:
 \begin{equation}
 \{\chi^{\underline{\alpha}}\}=\{\chi=-\tilde{a},\,\tilde{\zeta}_i\}\,,\label{axioncollective}
 \end{equation}
 where $\underline{\alpha}=0,\dots, k-1$.
 The quantization of the axion-charges follows from the identifications:
 \be \label{shiftaxion}
\chi^{\underline{\alpha}}=\chi^{\underline{\alpha}}+L_{\underline{\alpha}}\,\,\leftrightarrow\,\,\,\tilde{a} = \tilde{a} - L_0\,,\quad \tilde{\zeta}_i= \tilde{\zeta}_i + L_i\,,
\ee
 where $L_{\underline{\alpha}}=L_0,\,L_i$ are the radii of the axion-circles. In light of the precise identification of the fields $\chi^{\underline{\alpha}}$ with the corresponding string excitations, to be discussed below, we shall eventually set $L_{\underline{\alpha}}=1$.

 This quantisation was carried in \cite{Ruggeri:2017grz}
 and can be stated as the following condition:
\begin{equation}\label{quantcond}
2\pi\,\frac{n_{\underline{\alpha}}}{L_{\underline{\alpha}}}=\int_{S^4} dS^\mu\,J_{\underline{\alpha}|\mu}={\rm Vol}(S^4)\,J_{\underline{\alpha}|\tau}\,\,,\,\,\,\,n_{\underline{\alpha}}\in \mathbb{Z}\,,
\end{equation}
where $J_{\underline{\alpha}|\mu}$ are the Noether charges associated with the axion shift-symmetries, see equation (\ref{NC}) of Appendix \ref{Solvpar}.
The explicit expression of $n_{\underline{\alpha}}$ in our solutions is \footnote{This corrects a missing factor of $1/2$ in \cite{Ruggeri:2017grz}. Moreover, since we are considering here the axion $\chi_{0}=-\tilde{a}$, which corresponds to the ten dimensional type IIB axion field $\chi$, there is a sign difference between in the expression of $n_0$ with respect to the same reference.  }:
\begin{align}\label{eq:charge-quant}
& n_0 = -e^{-2U(0)}(m_0 - p_0) \frac{\rm Vol(S^4)}{2\kappa_5^2}\frac{L_0}{2\pi} \,,\nonumber\\
& n_i = \left( e^{-U(0)}(m_i - p_i) + e^{-2U(0)}(m_0 - p_0)\zeta^i(0)\right)\frac{ \rm Vol(S^4)}{2\kappa_5^2}\frac{L_i}{2 \pi} \,,
\end{align}
with $n_0, n_i \in \mathbb{Z}$.
The actual values of the circle lengths can be derived from the string theory origin of all fields. Instead of pursuing that option, we simply carry the $L$'s around in our expressions and only fix them when we compute certain dual gauge theory quantities.\\
We will find it convenient, when discussing holography, to introduce constant quantities $n'_{\underline{a}}$, $\underline{a}=0,\dots, k-1$, associated with the non axionic scalar fields. To this end let us recombine the non-axionic scalar fields $U,\, \zeta^i$ in the following way:
\begin{equation}
\varphi^{\underline{a}}\equiv\,\{\delta,\,\zeta^i\}\,,
\end{equation}
where the new scalar $\delta$ is defined as follows:
\begin{equation}
\delta\equiv e^{2U}+\frac{1}{2}\sum_{i=1}^{k-1} (\zeta^i)^2\,.
\end{equation}
We define the quantities $n'_{\underline{a}}$ on our solutions as follows:
\begin{equation}
n'_{\underline{a}}=\frac{{\rm Vol}(S^4)L_{\underline{a}}}{2\pi}\,K^{-1}{}_{{\underline{a}}}{}^{\mathcal{I}}(\phi_0)
\,
J_{\mathcal{I}|\tau}(\phi_0)=\frac{{\rm Vol}(S^4)L_{\underline{a}}}{2\pi\,2\kappa_5^2}\,
\,
G_{\underline{a}\mathcal{K}}(\phi_0)\,\left.\partial_\tau \phi^{\mathcal{K}} \right\vert_{\tau=0}\,,\label{naxch}
\end{equation}
where $K^{-1}{}_{{\mathcal{I}}}{}^{\mathcal{J}}$ is the inverse matrix of $K_{{\mathcal{I}}}{}^{\mathcal{J}}$ which describes the components of the Killing vectors $K_{{\mathcal{I}}}$ defined in Appendix \ref{Solvpar}. Their explicit expression is
\begin{align}\label{nonaxch}
& n'_0 = -e^{-2U(0)}(m_0 + p_0) \frac{\rm Vol(S^4)}{2\kappa_5^2}\frac{L_0}{2\pi} \,,\nonumber\\
& n'_i = \left( e^{-U(0)}(m_i + p_i) + e^{-2U(0)}(m_0 + p_0)\zeta^i(0)\right)\frac{ \rm Vol(S^4)}{2\kappa_5^2}\frac{L_i}{2 \pi} \,.
\end{align}
Note that the expression of $n'_{\underline{a}}$ is obtained from  that of $n_{\underline{\alpha}}$ by changing $p_\alpha\rightarrow -p_\alpha$. As opposed to the latter, the former in general are not integers. Note that we have used, for notational convenience, two different indices $\underline{\alpha}$ and $\underline{a}$, with the same range of values, in order to label the axionic and non-axionic scalar fields, respectively. Later, however, we will find it convenient to use a single index $\underline{\alpha}$ to label the two kinds of scalar fields, when the interpretation of the corresponding quantities is clear from the context.

\subsection{On-shell actions}

We now present the results of \cite{Ruggeri:2017grz} for the the Euclidean on-shell action, which is generically complex.
The imaginary part of the on-shell supergravity action, for all instantons, is particularly simple and only a function of the quantised charges (and not the number $c$):
\begin{equation}\label{im-action}
S^{\rm im} =  2\pi\,\Bigl(i \tilde{a}(0)\, \frac{n_0}{L_0} + i\sum_j\tilde{\zeta}_j(0)\, \frac{n_{j}}{L_j}\,\Bigr)\,.
\end{equation}
The corresponding expression for the real part of the on-shell action is
\begin{equation}\label{useful}
S^{\rm real} =  2\pi\,\Bigl(\left[\tilde{a}(\infty)-\tilde{a}(0)\right]\, \frac{n_0}{L_0} + \sum_j\left[\tilde{\zeta}_j(\infty)- \tilde{\zeta}_j(0)\right]\, \frac{n_j}{L_j}\,\Bigr)\,.
\end{equation}

An explicit computation for the general (sub-)extremal instanton, i.e. excluding wormholes, gives
\begin{equation}
S^{\rm real}=\frac{\text{Vol}(S^4)}{2\,\kappa^2_5}\,\frac{1}{\hat{m}^2_0\hat{p}^2_0 }\,{\rm Abs}\left[\frac{(\hat{m}_0+\hat{p}_0)}{2}\,\sum_{i=1}^{k-1}\,\left(\hat{m}_0\,p_i-\hat{p}_0\,m_i\right)^2- \mu\,\hat{m}_0\hat{p}_0(\hat{m}_0-\hat{p}_0)^2 \right]\,,\label{Sbdryne}
\end{equation}
where we have defined:
\begin{equation}
\hat{m}_0=m_0-\mu\,,\,\,\,\,\hat{p}_0=p_0-\mu\,\,,\,\,\,\,\,\mu=\sqrt{\vec{m}\cdot\vec{p}}\,.
\end{equation}
For regular wormholes the on-shell action is the difference between the left and right hand side of the boundary values and we refer to \cite{Bergshoeff:2005zf, Hertog:2017owm} for some explicit expressions when $k=1$.
In the extremal limit $\mu\rightarrow 0$, the expression \eqref{Sbdryne} simplifies drastically, and reads
\begin{equation}
S^{{\rm real}}_{{\rm SUSY}}=\frac{\text{Vol}(S^4)}{2\,\kappa^2_5}\,\left[|m_0+p_0|\,\left(1 + \frac{1}{2}\sum_{i=1}^{k-1}\Bigr[\frac{m_i^2}{m_0^2}+ \frac{p_i^2}{p_0^2}\Bigl]\right)\right]\,,\label{boundintegralExtr}
\end{equation}
where the extremality constraint $\vec{m}\cdot\vec{p}=m_0 p_0+\sum_{i=1}^{k-1}m_i p_i=0$ is understood to be imposed.

The final results required for the holographic match in the next sections are the asymptotic expansions for the scalar fields near the boundary. For that purpose, the following choice of coordinates is useful
\begin{align}\label{eq:new-vars-BPS}
\tilde{\phi}^{\mathcal{I}}=\{\delta,\,\zeta^i,\,\tilde{\zeta}_i,\,-\tilde{a}\}\,,
\end{align}
where $\delta$ and $\tilde{a}$ were defined earlier,
and we have used the tilde symbol to distinguish these new scalars from the old ones $\phi^{\mathcal{I}}=\{U,\,\zeta^i,\,\tilde{\zeta}_i,\,\tilde{a}\}$. The explicit metric on the moduli space in the $\tilde{\phi}$ basis is:
\begin{align}\label{eq:new-metr-BPS}
\tilde{G}_{\mathcal{IJ}}=e^{-4U}\left(\begin{matrix} 1 & - \, \zeta^i & 0 & 0 \cr -\, \zeta^i & \,(e^{2U}\delta^{ij} + \zeta^i  \zeta^j) & 0 & 0 \cr 0 & 0  &-\,(e^{2U}\delta^{ij} + \zeta^i  \zeta^j) &  \zeta^i\cr
0 & 0 &  \zeta^i & -1
\end{matrix}\right)\,,
\end{align}
which will be used explicitly in due course.
%

\section{Instantons in quiver gauge theories}\label{sec:instantons}
The understanding of instantons in gauge theories has a long history with interesting spin-offs in pure mathematics and we refer the reader to the excellent reviews \cite{Dorey:2002ik, Tong:2005un , Vandoren:2008xg} which are of particular relevance for the discussions here.

Since we only describe specific aspects of the holographic dual to the geodesic curves we do not need many details of instanton solutions and in what follows we therefore present the minimal information required for this paper, such as the one-point functions, Pontryagin indices and on-shell actions.
If we furthermore ignore the bi-fundamental matter fields and the fermions in the necklace quiver gauge theory then the classical action truncates simply into $k$ decoupled $\SU(N)$ pure Yang-Mills theories:\footnote{We use the following notation: $$\text{Tr}[F^2]\equiv \text{Tr}[F_{\mu\nu}\,F^{\mu\nu}]\,\,,\,\,\,\text{Tr}[F\wedge F]\equiv \text{Tr}[F_{\mu\nu} \star_4 F^{\mu\nu}]\,,$$
where, in flat space-time, $\star_4 F_{\mu\nu}\equiv \frac{1}{2}\,\epsilon_{\mu\nu\rho\sigma}\,F^{\rho\sigma}$.}
\be
\mathcal{L} = \sum_{{\alpha}=0}^{k-1} \left( -\frac{1}{4g_\alpha^2}\text{Tr}[F_{\alpha}^2] - i\frac{\theta_{\alpha}}{32\pi^2}\,\text{Tr}[F_{\alpha}\wedge F_{\alpha}] \right) \,.\label{actionFF}
\ee
Our notation is such that $F_{\alpha}$ is the field strength in the ${\alpha}$'th $\SU(N)$ gauge factor (node). The factor $i$ in front of the theta-angles arises because we are working in Euclidean signature. Our conventions for the Lie algebra valued fields are:
\be
\text{Tr}[T_aT_b]=-\delta_{ab}\,,\qquad F^a_{\mu \nu}=\partial_{\mu} A^a_{\nu} - \partial_{\nu} A^a_{\mu} + f^a_{bc}A^b_{\mu}A^c_{\nu}\,,
\ee
where $a, b \dots$ denote $\SU(N)$ Lie algebra indices. It could very well be that the above truncation is too simplistic to capture all instantons, but it will suffice for most of the discussion in this paper.

We are aware of three classes of instanton solutions that all rely on self-duality in certain sectors of the product gauge group $\SU(N)^k$.  The moduli-spaces of these instantons are quite involved but we will mostly focus on their Pontryagin labels:
\be
N_{\alpha} = -\frac{1}{32\pi^2}\int_4 \d^4 x \, \text{Tr}[F_{\alpha}\wedge F_{\alpha}]\,.
\ee
Therefore an important characterisation of classes of instanton solutions in quiver gauge theories is to provide a string of $k$ integers:
\be
(N_0, \ldots, N_{k-1})\,.
\ee
We shall also define the following non-negative quantities:
\be
N'_{\alpha} = -\frac{1}{32\pi^2}\int_4 \d^4 x \, \text{Tr}[F_{\alpha}^2]\ge 0\,.
\ee
As opposed to the $N_\alpha$, the $N'_\alpha$ are in general not quantized.
\subsubsection*{Extremal instantons}

The supersymmetric instantons of the quiver theory have the property that each field strength obeys (anti-) self duality:
\be \label{selfdual}
\star_4 F_{\alpha} = \pm F_{\alpha}\,,
\ee
and furthermore all field strengths have the same orientation. This means that all $N$'s are of the same sign. So either all field strengths $F_{\alpha}$ are self dual or all are anti-self dual. In other words there is no mixture of instantons and anti-instantons. These configurations preserve 4 real supercharges.

If one allows for mixtures of instantons and anti-instantons, the solutions still solve the equations of motion because each gauge node separately obeys (\ref{selfdual}). These configurations were coined ``quasi-instantons" in the literature, see for instance \cite{Imaanpur:2008jd}, and also correspond to local minima of the action, but necessarily break supersymmetry. Their importance in supersymmetric gauge theories seems unclear.
For all the above extremal (SUSY and quasi-instanton) solutions the following equality holds:
\begin{equation}
N'_\alpha=|N_\alpha|\,,
\end{equation}
and the on-shell action is given by:
\be\label{inst-action}
S = \sum_{\alpha} \left( \frac{8\pi^2}{g_{\alpha}^2}|N_{\alpha}|+ i \theta_{\alpha} N_{\alpha} \right) \,,
\ee
while the expressions for the classical gauge of single centered configurations read
\be \label{one-point}
\text{Tr}[F^2_{\alpha}]=\text{sign}(N_{\alpha}) \text{Tr}[F_{\alpha}\wedge F_{\alpha}] = -192\frac{z_0^4}{\left(z_0^2 + (\vec{x}-\vec{x}_0)^2\right)^4}|N_{\alpha}|\,.
\ee
Here, $\vec{x}_0$ is the position of the instanton, which we have taken to be the same for all gauge nodes for simplicity, and $z_0$ sets the thickness radius of the instanton, again taken to be the same for all gauge nodes.

\subsubsection*{Non-extremal instantons}
We consider the above two classes of instantons as ``extremal" because they satisfy (anti-)self-duality in each separate gauge node. However there also exist classical YM solutions with finite action that are not (anti-) self-dual. A particularly simple class of such solutions can be constructed in theories with large $N$ as pointed out in \cite{Bergshoeff:2005zf}. The idea is rather straightforward.  One can consider mutually commuting $\SU(2)$-factors inside $\SU(N)$. If one then takes the separate  $\SU(2)$ to be (anti-) self dual then such a gauge field configuration solves the equations of motion without obeying (\ref{selfdual}). For the sake of illustration we take a single anti-instanton configuration for $\SU(2)$ denoted, $\overline{A}_{\mu}$, inside the color matrix which for the rest has $\SU(2)$ instantons on the diagonal:
\be
A_{\mu}^{\rm SU(N)}=\begin{pmatrix}
A_{\mu}^{\rm SU(2)} & 0 & \ldots & 0\\
 0 & A_{\mu}^{\rm SU(2)} &  & 0\\
 \vdots  &  & \ddots  & \\
0   & &  & \overline{A}_{\mu}^{\rm SU(2)}
\end{pmatrix}\,.
\ee
If we call $N^+_{\alpha}$ the number of $\SU(2)$-instantons and $N^-_{\alpha}$ the number of anti-$\SU(2)$-instantons inside the ${\alpha}$'th gauge node then the on-shell action and gauge field profiles are given by
\begin{align}
& S = \sum_{\alpha} \frac{8\pi^2}{g_{\alpha}^2}(N_{\alpha}^+ + N_{\alpha}^-) + i \theta_{\alpha} (N_{\alpha}^+ - N_{\alpha}^-)\,,\\
& \text{Tr}[F^2_{\alpha}]= -192\frac{z_0^4}{\left(z_0^2 + (\vec{x}-\vec{x}_0)^2\right)^4}(N_{\alpha}^+ + N_{\alpha}^-)\,,\\
& \text{Tr}[F_{\alpha}\wedge F_{\alpha}]= -192\frac{z_0^4}{\left(z_0^2 + (\vec{x}-\vec{x}_0)^2\right)^4}(N_{\alpha}^+ - N_{\alpha}^-)\,,
\end{align}
where $\vec{x}_0$ and $z_0$ are again the common position and equal thickness of the instantons.
It should be obvious from the above that the Pontryagin indices for the $\alpha$-th gauge node is by definition
\be
N_{\alpha} = (N_{\alpha}^+ - N_{\alpha}^-)\,,
\ee
and as a consequence the imaginary part of the on-shell action, as well as the Tr$[F_{\alpha}\wedge F_{\alpha}]$ expression is the same for all instanton solutions considered above.

\section{Holographic correspondence}\label{sec:hol-map}
In this section we describe the match between the on-shell actions on both sides of the correspondence and the dual one-point functions of  $\text{Tr}[F_{\alpha}^2]$ and $\text{Tr}[F_{\alpha}\wedge F_{\alpha}]$.

One obvious correspondence should be the exact map between the topological data on both sides of the correspondence: the Pontryagin indices $N_{\alpha}$ and the quantised axion charges $n_{\underline{\alpha}}$. The most general relation between the two strings of integers is
\begin{equation}\label{correspondencen}
N_{{\beta}}=\sum_{\underline{\alpha}=0}^{k-1} n_{{\underline{\alpha}}}\,M_{\underline{\alpha}\beta}\,,
\end{equation}
with $M \in \GL(k,\mathbb{Z})$. Here we have distinguished between the index $\alpha=0,\dots, k-1$, which labels the vertices of the quiver diagram, from $\underline{\alpha}$ labeling the quantized axion charges of our solution. For later purposes it is useful to remark that the quiver diagram exhibits the cyclic symmetry $\mathbb{Z}_k$ of the corresponding extended Dynkin diagram $A_{k-1}^{(1)}$:
\begin{align}
\mathbb{Z}_k&=\{\boldsymbol{\nu}^\ell\}_{\ell=0,\dots,k-1}\,\,,\,\,\,\,\,\,\,\,\,\,\,
\boldsymbol{\nu}\,:\,\begin{cases}\alpha_0\rightarrow \alpha_1\cr \alpha_{1}\rightarrow \alpha_{2}\cr \vdots \cr\alpha_{k-1}\rightarrow \alpha_{0}\end{cases}
\end{align}
where $\alpha_0,\,\dots,\,\alpha_{k-1}$ are the $k$ simple roots of $A_{k-1}^{(1)}$.

\subsection{Coordinates on moduli space versus conformal manifold}
In order to study a possible holographic correspondence between the supergravity instantons and field theory instantons we need the detailed dictionary between the moduli and the YM couplings $g_{\alpha}, \theta_{\alpha}$. To the best of our knowledge this is not known.  Our working assumption is therefore that, like in AdS$_5\times$ S$^5$ we might hope for a match between the on-shell actions of the \emph{supersymmetric} instantons:
\begin{equation}
S_{\text{gauge theory}}(N_{\alpha}) = S_{\text{SUGRA}}(n_{\underline{\alpha}})\,\,\,.
\end{equation}
This equality constraints the dictionary between the coordinates on the conformal manifold and the coordinates on the AdS moduli space. Once we fix from this the dictionary, we compute, as a consistency test, the dual one-point functions for the operators  $\text{Tr}[F_{\alpha}^2]$ and $\text{Tr}[F_{\alpha}\wedge F_{\alpha}]$. These one-point functions should be consistent with the gauge theory expectations and this will turn out to be the case. Then, armed with the correct dictionary, we can work out the consequences for the non-SUSY instantons.

What furthermore helps us is to have a 10-dimensional picture of the type IIB supergravity moduli dual to the conformal manifold. There is the 10d axio-dilaton $u=\chi+i\,e^{-\phi}$ and there are the vevs of the $B_2$ and $C_2$ field over the collapsing two-cycles $\Sigma_I$ of the orbifold:
\begin{equation}
v_I =-\int_{\Sigma_I}(B^{{\rm RR}}_{(2)}-u \,B^{{\rm NS}}_{(2)})\,,\label{vBB}
\end{equation}
where $I=1\ldots k-1$.

At this point we have three sets of scalar fields: from 10d SUGRA we have $k$ complex fields
\begin{equation}
\text{10D supergravity:}\quad t_{\underline{\alpha}}=\{u,\,v_I\}\,,
\end{equation}
from 5d SUGRA we have $2k$ real fields
\begin{equation}
\text{5D supergravity:}\quad U,a, \zeta^i, \tilde{\zeta}_i\,,
\end{equation}
and we have $k$ complexified gauge couplings
\begin{equation}
\text{4D gauge theory:}\quad t_{\alpha}= \frac{\theta_{\alpha}}{2\pi} + i \frac{4\pi}{g^2_{\alpha}}\,.
\end{equation}
We now relate all of them.

First we note that the action of quiver-symmetry $\mathbb{Z}_k$ on the complexified coupling constants $t_\alpha$ (and the corresponding field strengths $F^\alpha$) leaves the classical field theory action (\ref{actionFF}) invariant. This symmetry group $\mathbb{Z}_k$ should be contained in the compact part ${\rm O}(k)$ of the isotropy group ${\rm GL}(k,\mathbb{R})$ of the moduli space and should also act \emph{linearly} on the scalar fields. One can show that only the subgroup ${\rm O}(k-1)$ of ${\rm O}(k)$ has a linear action on the scalar fields and it only acts on $\zeta^I,\,\tilde{\zeta}_I$ leaving $a$ and $U$ invariant. Therefore $a, U$ have to correspond to the complex axio-dilaton field $u$, which is a $\mathbb{Z}_k$-singlet since it is insensitive to the orbifold geometry. Similarly the $\zeta^I,\,\tilde{\zeta}_I$ should be related to the vevs of the NS-NS and R-R 2-forms $B^{{\rm NS}}_{(2)},\,B^{{\rm RR}}_{(2)}$ across the shrinking 2-cycles $\Sigma_I$ of the orbifold. So purely based on the action of the $Z_k$ symmetry we can deduce
\begin{align}
u & = \chi+i\,e^{-\phi}\sim -\tilde{a}+i\,\delta\,,\nonumber\\
v_I & \sim \tilde{\zeta}_I+ i \zeta_I \,. \label{identi}
\end{align}
We choose our parametrization so that the proportionality constants are fixed to one:
 \begin{equation}
v^I=\tilde{\zeta}_I+ i \zeta_I\,,\qquad u = -\tilde{a}+i\,\delta\,.
\end{equation}
Although this amounts to choosing the parameters $L_{\underline{\alpha}}$ to be unity, for convenience we shall keep those periods in the formulae below.\footnote{We shall recover this identification in Appendix \ref{App:IIB}, though through a different analysis which consists in  comparing the moduli spaces of two different theories: Type IIB on $AdS_5\times S^5/\mathbb{Z}_2$ and on $AdS_5\times T^{1,1}$. The two internal manifolds share the same topology although the former is singular while the latter can be viewed as a smoothed version of it. In principle the identification of the moduli and the geometry of the moduli space only depend on the topology. However the singular nature of $S^5/\mathbb{Z}_2$, as opposed to $T^{1,1}$, implies that, while the metrics on the moduli spaces match to lowest order in $|v^I|$, confirming the above identification, the two differ by terms which are of higher order in $|v^I|$. We shall discuss this point in more detail in the Appendix.}
The $v^I$ define the \emph{twisted moduli} and are in one-to-one correspondence with the simple roots $\alpha_I$, $I=1\dots, k-1$, of the $A_{k-1}$ algebra. Their non-trivial transformation property under the $\mathbb{Z}_k$ symmetry of the quiver diagram, to be described in detail below,  differentiates them from the \emph{untwisted complex scalar} $u$ .  \footnote{We distinguish here the index $I$ from the index $i=1\dots, k-1$ which labels an orthonormal basis of the Cartan subalgebra of $A_{k-1}$. The relation between the two bases is encoded in a matrix ${\bf C}=(C_i{}^I)$:
$$\mathcal{A}^{-1|IJ}\,=\,\sum_i\,C_i{}^IC_i{}^J\,,$$
where $\mathcal{A}_{IJ}$ is the Cartan matrix of $A_{k-1}$. We will therefore have $v_i= C_i{}^I\,v_I$, $m_i=C_i{}^I\,m_I$ and $p_i=C_i{}^I\,p_I$.}
Then,  from the discussions in references \cite{Corrado:2002wx,Nekrasov:2012xe} we can relate
 $t_\alpha=\{t_0,\,t_I\}$ to  $t_{\underline{\alpha}}=\{u,\,v_I\}$ as follows:
\begin{equation}
\frac{t_{\underline{\alpha}}}{L_{\underline{\alpha}}}=\sum_\alpha M_{\underline{\alpha}\alpha}\,t_\alpha\,,\label{tauaa}
\end{equation}\
such that\footnote{Note that the matrix $M$ has the following property:
\begin{equation}
\sum_{\alpha=0}^{k-1}M_{\underline{\alpha}\alpha}M_{\underline{\beta}\alpha}=\left(\begin{matrix}k & {\bf 0}_J \cr {\bf 0}_I & \mathcal{A}_{IJ}\end{matrix}\right)\,,
\end{equation}
where $\mathcal{A}_{IJ}$ is the Cartan matrix of $A_{k-1}$.}
\begin{equation}
u =\sum_{\alpha=0}^{k-1} t_\alpha\,\,,\qquad
v_I\,=\,-(t_{I-1}-t_{I})\,\,,\,\,\,\,I=1,\dots, k-1\,\,.\label{identi0}
\end{equation}
From the above equations we can determine the form of the matrix $M$.
For instance in the $k=2$ case we find:
\begin{equation}
M_{\underline{\alpha}\alpha}=\left(\begin{matrix}1 & 1\cr -1 & 1\end{matrix}\right)\,\,\Rightarrow\,\,\,(M^{-1})_{\alpha\underline{\alpha}}=\frac{1}{2}\left(\begin{matrix}1 & -1\cr 1 & 1\end{matrix}\right)\,.\label{matrixM}
\end{equation}

We now investigate whether the on-shell actions on both sides of the duality are equal. From the on-shell imaginary action on the supergravity side is given by \eqref{im-action}. This expression is the same for all instantons, SUSY or not and captures purely topological information.
A useful rewriting of the imaginary part is given by:
\be
S_{\text{GRAV}}^{Im} = \sum_{\underline{\alpha}}\frac{2\pi}{L_{\underline{\alpha}}}\left[i\,{\rm Re}(t_{\underline{\alpha}})\,n_{\underline{\alpha}})\right]  \,.\label{osactionnewIm}
\ee
Similarly, a useful shorthand way to rewrite the real part of the on-shell supergravity action is
\be
S^{\text{Real}}_{\text{GRAV}} = \sum_{\underline{\alpha}}\frac{2\pi}{L_{\underline{\alpha}}}\left[{\rm Im}(t_{\underline{\alpha}})\,n'_{\underline{\alpha}}\right] +\text{constant} \,,\label{osactionnewreal}
\ee
where $t_{\underline{\alpha}}$ is given by
\begin{equation}
t_{\underline{0}} = u =- \tilde{a}(0) + i \delta(0)\,,\qquad t_{\underline{I}} = \tilde{\zeta}_I(0) +i \zeta^I(0)\,,
\end{equation}
and the constant term refers to a contribution that does not depend on the moduli at the boundary. More explicitly we have:
\be
S^{\text{Real}}_{\text{GRAV}} = \sum_{\underline{\alpha}}\frac{2\pi}{L_{\underline{\alpha}}}\left[{\rm Im}(t_{\underline{\alpha}}(0)-t_{\underline{\alpha}}(\infty))\,n'_{\underline{\alpha}}\right] \,.
\ee
For the supersymmetric solutions the $n'_{\underline{\alpha}}$ are, up to a possible minus signs, equal to the $n_{\underline{\alpha}}$ and so the real part of the action is also a weighted sum of integers, just like the on-shell YM action. If one insists on equating the on-shell actions on both sides of the duality one would conclude that the gauge couplings are dual to linear combinations of the scalar vevs in the UV subtracted by the values at $\tau=\infty$ (which means $z=\ell_{AdS}$). For AdS$_5\times S^5$ this issue does not arise since there ${\rm Im}(t_0(\infty))=0$ always. For the sake of computing holographic $n$-point functions the distinction between identifying the dual couplings as the UV values or the UV values subtracted with the values in the bulk at $\tau=\infty$, does not matter since variations of the action with respect to $t_{\underline{\alpha}}(0)$ or $t_{\underline{\alpha}}(0)-t_{\underline{\alpha}}(\infty)$ are equal. Hence for notational simplicity we disregard now the $t_{\underline{\alpha}}(\infty)$ contribution. As mentioned in the Introduction, understanding this constant mismatch in the context of AdS/CFT will be object of a future investigation.

Now, we are ready to equate the on-shell action of the supersymmetric instantons on both sides of the holographic correspondence. The on-shell YM action is
\be
S_{\text{GAUGE}} = \sum_{{\alpha}}{2\pi}\left[{\rm Im}(t_{{\alpha}})\,|N_{{\alpha}}|+i\,{\rm Re}(t_{{\alpha}})\,N_{{\alpha}})\right] \,.\label{gauge action}
\ee
In view of \eqref{tauaa}, the match works if we identify:
\begin{align}
n'_{\underline{\alpha}}=M^{-1}_{\alpha\underline{\alpha}}\,|N_{\alpha}|\,\,,\,\,\,\,\,
n_{\underline{\alpha}}=M^{-1}_{\alpha\underline{\alpha}}\,N_{\alpha}\,,\label{nN}
\end{align}
where we can use
\be
\sum_{\underline{\beta}\gamma}M_{\underline{\beta}\alpha}M^{-1}_{\gamma\underline{\beta}}|N_{\gamma}| = |N_{\alpha}|\,.
\ee
We can generalize the relations (\ref{nN}) to the non-extremal case defining:
\begin{equation}
N'_\alpha=-\frac{1}{32\pi^2}\int {\rm Tr}[F_\alpha^2]\ge 0\,,
\end{equation}
and writing:
\begin{align}
n'_{\underline{\alpha}}=M^{-1}_{\alpha\underline{\alpha}}\,N'_{\alpha}\,\,,\,\,\,\,\,
n_{\underline{\alpha}}=M^{-1}_{\alpha\underline{\alpha}}\,N_{\alpha}\,.\label{nNnE}
\end{align}
In the extremal case in which $N'_{\alpha}=|N_{\alpha}|$ we recover equations (\ref{nN}).
In this section, with an abuse of notation, we have denoted by $n_{\underline{\alpha}},\,n'_{\underline{\alpha}}$ quantities depending on the vector field strengths at the boundary, while we have used the same symbols in section \ref{acq} to denote background quantities associated with a supergravity solution. We shall prove in what follows that the vevs of the former quantities, computed using the holographic correspondence, coincide indeed with the latter denoted by the same symbols.

\subsection{Holographic one-point functions}
Consider the background described by the geodesic $\phi^{\mathcal{I}}(\tau,\,\phi_0)$. In the AdS/CFT correspondence \cite{Maldacena:1997re,Gubser:1998bc,Witten:1998qj} $\phi^{\mathcal{I}}_0$ are sources ${\rm J}^{\mathcal{I}}$ of dual operators $\mathcal{O}_{\mathcal{I}}$ in the CFT so that:
\begin{equation}
\langle \mathcal{O}_{\mathcal{I}}\rangle_{{\rm J}}=-\frac{\delta }{\delta {\rm J}^{\mathcal{I}}}S_{SUGRA}[{\rm J}]\,, \label{AdSCFT}
\end{equation}
where ${\rm J}^{\mathcal{I}}=\phi^{\mathcal{I}}_0$ and $S_{SUGRA}[{\rm J}]=S_{SUGRA}[\phi_0]$ is the supergravity action computed on the solution as a function of the boundary values of the scalar fields.
Computing the on-shell variation of $S_{SUGRA}$ as $\phi^{\mathcal{I}}_0\rightarrow \phi^{\mathcal{I}}_0+\delta \phi^{\mathcal{I}}_0$ one ends up with the following boundary term:
\begin{equation}
 \delta S=\frac{1}{2\kappa_5^2}\,\int_{\partial \mathcal{M}_5} \delta\phi^{\mathcal{I}}\,G_{\mathcal{IJ}}\,\partial_\mu\phi^{\mathcal{J}}d\Sigma^\mu =\int_{\partial \mathcal{M}_5}\delta\phi^{\mathcal{I}}_0\,\left(K^{-1}{}_{\mathcal{I}}{}^{\mathcal{J}}(\phi_0)
 \,J_{\mathcal{J}|\tau}\right)\,\partial_z \tau\,\,g_5^{zz}\,\sqrt{|g_5|}\,d^4x\,,\label{boundint}
 \end{equation}
where $J_{\mathcal{J}|\tau}$ are constants proportional to the Noether charges associated with the solvable isometries $T_{\mathcal{I}}$ and $d\Sigma^\mu$ the volume element on a Minkowski slice of the Euclidean asymptotically ${\rm AdS}$ space $\mathcal{M}_5$ at constant $z$.
Let us also use the fact that, near $z=0$, $g_5^{zz}\,\sqrt{|g_5|}\sim\ell^3/z^3$.
In the end we find:
\begin{equation}
\frac{\delta S}{\delta \phi_0^{\mathcal{I}}}=\lim_{z\rightarrow 0}\, \frac{\ell^3}{z^3}\,\partial_z \tau\,\left( K^{-1}{}_{\mathcal{I}}{}^{\mathcal{J}}(\phi_0)\,J_{\mathcal{J}|\tau}\right)\,.
\end{equation}
For the sake of simplicity, we localize the instanton at $\vec{x}_0=\vec{0}$ and $z_0=\ell$. As far as the axion fields $\chi^{\underline{\alpha}}$ are concerned, $K_{\underline{\alpha}}=\frac{\partial}{\partial \chi^{\underline{\alpha}}}$ and, using the property:
\begin{equation}
\lim_{z\rightarrow 0}\frac{\ell^3}{z^3}\,\partial_z\tau=\frac{16\, \ell^4}{[|\vec{x}|^2+\ell^2]^4}\,,
\end{equation}
we find:
\begin{equation}
\frac{\delta S}{\delta \chi^{\underline{\alpha}}}\,=\,\frac{16\, \ell^4}{[|\vec{x}|^2+\ell^2]^4}\,J_{\underline{\alpha}|\tau}=32\pi
\frac{n_{\underline{\alpha}}}{{\rm Vol}(S^4)\,L_{\underline{\alpha}}}\,\frac{ \ell^4}{[|\vec{x}|^2+\ell^2]^4}\,,
\end{equation}
where we have used the quantization condition (\ref{quantcond}) to express $J_{\mathcal{J}|\tau}$.
According to our previous analysis, the marginal operator dual to $\chi^{\underline{\alpha}}$ is:
\begin{equation}
\mathcal{O}_{\underline{\alpha}}=\frac{1}{16 \pi\,L_{\underline{\alpha}}}\,\sum_{\beta=0}^{k-1}M^{-1}{}_{\beta\underline{\alpha}}\,{\rm Tr}\left(F_\beta\wedge F_\beta\right)\,,
\end{equation}
such that:
\begin{equation}
\frac{1}{16 \pi\,L_{\underline{\alpha}}}\,\sum_{\beta=0}^{k-1}M^{-1}{}_{\beta\underline{\alpha}}\,\langle {\rm Tr}\left(F_\beta\wedge F_\beta\right)\rangle=-\frac{\delta S}{\delta \chi^{\underline{\alpha}}}\,=\,-32\pi
\frac{n_{\underline{\alpha}}}{{\rm Vol}(S^4)\,L_{\underline{\alpha}}}\,\frac{ \ell^4}{[|\vec{x}|^2+\ell^2]^4}\,.
\end{equation}
Using ${\rm Vol}(S^4)=8 \pi^2/3$ we end up with the following formula:
\begin{align}
\langle{\rm Tr}\left(F_\beta\wedge F_\beta\right)\rangle\,=\,-192\,\left(\sum_{\underline{\alpha}=0}^{k-1}M_{\underline{\alpha}\beta}
n_{\underline{\alpha}}\right)\,\frac{ \ell^4}{[|\vec{x}|^2+\ell^2]^4}\,=\,-192\,N_\beta\,\frac{ \ell^4}{[|\vec{x}|^2+\ell^2]^4}\,.
\end{align}
We can make a similar computation for the 1-point functions associated with the non-axionic scalars $\{\varphi^{\underline{a}}\}=\{\delta=e^\phi,\,\zeta^i\}$. The corresponding 1-point functions read
\begin{align}
\langle O_{\underline{a}}\rangle=\frac{1}{16 \pi\,L_{\underline{a}}}\,\sum_{\beta=0}^{k-1}M^{-1}{}_{\beta\underline{a}}\,\langle{\rm Tr}\left(F_\beta^2\right)\rangle=-\frac{\delta S}{\delta \varphi^{\underline{a}}}\,=\,- 32\pi
\frac{n'_{\underline{a}}}{{\rm Vol}(S^4)\,L_{\underline{a}}}\,\frac{ \ell^4}{[|\vec{x}|^2+\ell^2]^4}\,,
\end{align}
where we have used the definition of $n'_{\underline{a}}$ in (\ref{naxch}). By the same token we end up with the general formula:
 \begin{align}
\langle{\rm Tr}\left(F_\beta^2\right)\rangle\,=\,-192\,\left(\sum_{\underline{a}=0}^{k-1}M_{\underline{a}\beta}
n'_{\underline{a}}\right)\,\frac{ \ell^4}{[|\vec{x}|^2+\ell^2]^4}\,=\,-192\,N'_\beta\,\frac{ \ell^4}{[|\vec{x}|^2+\ell^2]^4}\,.
\end{align}
We find the relations between $N_\alpha,\,N'_\alpha$ and $n_{\underline{\alpha}},\,n'_{\underline{a}}$ given in the previous section:
\begin{align}
N'_\beta &\equiv  -\frac{1}{32 \pi^2}\,\int_{\mathcal{M}_4}\langle{\rm Tr}\left(F_\beta^2\right)\rangle=\sum_{\underline{\alpha}=0}^{k-1}M_{\underline{\alpha}\beta}
n'_{\underline{\alpha}}\,,\label{Npnp}\\
N_\beta &\equiv  -\frac{1}{32 \pi^2}\,\int_{\mathcal{M}_4}\langle{\rm Tr}\left(F_\beta\wedge  F_\beta\right)\rangle=\sum_{\underline{\alpha}=0}^{k-1}M_{\underline{\alpha}\beta}
n_{\underline{\alpha}}\,.\label{Nn}
\end{align}
\subsection{SUSY instantons}
For BPS solutions either $p_\alpha=0$ or $m_\alpha=0$, which implies
\begin{equation}
n'_{\underline{\alpha}}=\pm n_{\underline{\alpha}}\,,
\end{equation}
where the sign on the right hand side does not depend on $\underline{\alpha}$ and is $``+"$ if $p_\alpha=0$, $``-"$ otherwise. This implies:
\begin{equation}
N'_{\alpha}=\pm N_{\alpha}\,,
\end{equation}
which define BPS instantons in the dual theory. Note however that consistency requires $$N'_{\alpha}=|N_{\alpha}|\ge 0\,,$$
namely that $N'_{\alpha}$ be non-negative and quantized.
From the expression of the matrix $M$ and (\ref{Npnp}) we find:
\begin{align}
N'_0&=n'_0-n'_1\,,\nonumber\\
N'_1&=n'_0-n'_2+n'_1\,,\nonumber\\
&\vdots\nonumber\\
N'_{k-2}&=n'_{0}-n'_{k-1}+n'_{k-2}\,,\nonumber\\
N'_{k-1}&=n'_{0}+n'_{k-1}\,.
\end{align}
We need to require the right hand sides of the above equations (which are integers for the BPS solutions) to be non-negative. A sufficient condition for $k>2$ amounts to requiring:
\begin{equation}
(n'_0)^2\ge 2\,\sum_{i=1}^{k-1} (n'_i)^2\,,\label{conditionnp0}
\end{equation}
while for $k=2$ we only need $n'_0\ge |n'_1|$.
Note that $n'_0$, as defined in (\ref{naxch}), is always non-negative since regularity requires $m_0,p_0\le 0$. The above condition is a stronger one and is ${\rm O}(k-1)$-invariant. If we have a BPS solution, we can always rotate the charges so that (\ref{conditionnp0}) is satisfied, by means of a transformation in ${\rm O}(k)/{\rm O}(k-1)$. Once we are in this ${\rm O}(k-1)$-frame one can study the holographic correspondence.

\subsection{Intrinsically non-BPS solutions revisited}\label{revisited}
We now reconsider the intrinsically non-BPS solution discussed in subsection \ref{INBPS} for which $k=2$. We explicitly set now $L_0=L_1=1$  and define $\xi\equiv \frac{\rm Vol(S^4)}{2\kappa_5^2}\frac{1}{2\pi}$ and consider the geodesic moving through the origin. Recalling that regularity of the solution requires $m_0,\,p_0\,\le\, 0$, which implies ${\rm f}_0\le 0$, we find:
\begin{align}
&n'_0=-2\,{\rm f}_0\,\xi=2\,\xi\,|{\rm f}_0|\,\,,\,\,\,\,n'_1=0\,,\nonumber\\
&n_0=0\,\,,\,\,\,\,n_1=2\,{\rm f}_1\,\xi\,.\label{nnpmp}
\end{align}
The explicit expression of the matrix $M_{\underline{\alpha}\alpha}$ in (\ref{nNnE}) and (\ref{tauaa}) is given in (\ref{matrixM}), and we get:
\begin{align}
N_0'&=-2\,{\rm f}_0\,\xi=2\,\xi\,|{\rm f}_0|\,\,,\,\,\,N_1'=-2\,{\rm f}_0\,\xi=2\,\xi\,|{\rm f}_0|\,,\nonumber\\
N_0 &=-2{\rm f}_1\,\xi \,\,,\,\,\,N_1=2\,{\rm f}_1\,\xi\,,
\end{align}
where we recall that regularity of the solution requires ${\rm f}_0< 0$.
Note that in the super-extremal case, $c<0$, we have $|{\rm f}_0|<|{\rm f}_1|$ so that:
\begin{equation}
N_\alpha'< |N_\alpha|\,\,\Rightarrow\,\,\,\,\vert {\rm Tr}[F^2_\alpha]\vert\,<\, |{\rm Tr}[F_\alpha\wedge {F}_\alpha]|\,.
\end{equation}
We therefore find a violation of the positivity, \emph{aka} BPS bound in the dual gauge theory. Note that this result was already anticipated in \cite{Bergshoeff:2005zf} but could not be proven since the wormholes in \cite{Bergshoeff:2005zf} were not regular.  The fact that for $c <0$ we violate the BPS bound is consistent with the recently found instabilities (in the Euclidean sense) of macroscopic axion wormholes \cite{Hertog:2018kbz}. This means that the wormholes do not contribute in any path integral since there is a similar configuration with the same boundary conditions that has lower action. This configuration is expected to be the complete defragmentation of the macroscopic wormholes into a dilute gas of microscopic wormholes, each carrying a unit axion charge. Such a configuration has no classical description. The fact that we find the regular wormholes to violate the BPS bound in the dual gauge theory constitutes holographic evidence for the results in \cite{Hertog:2018kbz} that the macroscopic wormholes are indeed unphysical.

For the subextremal solutions, $c>0$, we have $|{\rm f}_0|>|{\rm f}_1|$ and we hence satisfy the BPS bound without saturating it since $N'_{\alpha}>|N_{\alpha}|$. Hence we find a correspondence to the non-extremal gauge theory instantons presented in the previous section.

In the extremal limit $|{\rm f}_0|=|{\rm f}_1|$, i.e. ${\rm f}_0=\pm{\rm f}_1$. This implies  $N_0'= |N_0|=\,,N_1'=|N_1|$ and, more precisely,  $N_0'=\pm N_0\,,N_1'=\mp N_1$. In terms of the field strengths of the boundary theory these relations imply
\begin{equation}
\vert {\rm Tr}[F^2_0]\vert\,=\, \mp {\rm Tr}[F_0\wedge {F}_0]\,\,,\,\,\,\vert {\rm Tr}[F^2_1]\vert\,=\, \pm {\rm Tr}[F_1\wedge {F}_1]\,,
\end{equation}
which is the characterization of a quasi-instanton. We conclude that the solution discussed in subsection \ref{INBPS} provides, in the extremal limit, the supergravity dual of a quasi-instanton. Hence the name `intrinsically non-BPS' since there is no continuous change of parameters that brings us to a supersymmetric solution. Interestingly this is where the smooth wormholes are residing in the super-extremal corner.

To summarise: our simple solution shows that, at the level one-point functions, we can holographically reproduce non-SUSY quasi instantons in quiver gauge theories as well as non-extremal Yang Mills solutions. The super-extremal gravity solutions, corresponding to smooth wormholes, on the other hand have no dual description consistent with the findings in \cite{Hertog:2018kbz} that they do not contribute to the quantum gravity path integral.

\section{Discussion}\label{sec:discussion}
We have investigated the holographic correspondence between supergravity instantons in $\rm AdS_5\times S^5/\mathbb{Z}_k$, and instantons in the  $\mathcal{N}=2$ necklace quiver gauge theory in four dimensions with $k$ nodes (labelled with $\alpha=0\ldots k-1$) at large N.  We have compared the on-shell actions and the  vevs of the operators Tr$[F_{\alpha}^2]$ and Tr$[F_{\alpha}\wedge F_{\alpha}]$ between instantons on the field theory and instantons on the supergravity side. We have found that the match between the supersymmetric instantons on both sides of the correspondence works out in detail. This also allowed us to write the precise dictionary between the supergravity scalars and the gauge couplings.

However there is more than just the supersymmetric saddle points. On the supergravity side 2 other classes of instantons were found in \cite{Ruggeri:2017grz} that break supersymmetry: 1) extremal ($c=0$) but non-supersymmetric solutions, and 2) non-extremal ($c\neq 0$) solutions. The gauge-theory seems to have the same property.  The dual to the non-SUSY but extremal instantons are gauge theory instantons for which different gauge theory nodes carry instantons of different orientation, which we named ``quasi-instantons'' as in \cite{Imaanpur:2008jd}. Our analysis based on the computation of 1-point functions provides evidence that the extremal non-supersymmetric solution discussed in section \ref{revisited} provide the supergravity dual to a ``quasi-instanton'' solution.

The dual to some of the non-extremal instantons with $c>0$ was already suggested in \cite{Bergshoeff:2005zf} and corresponds to taking some $\SU(2)$ sub-algebras inside a separate gauge node and flip the orientation of some of these mutually commuting $\SU(2)$ sub-blocks. The dual to wormhole solutions ($c<0$) cannot make sense since we have explicitly demonstrated that the dual one-point functions computed from everywhere smooth wormholes, violate the BPS bound in the dual field theory. This is a very sharp AdS/CFT paradox but gets resolved if one accepts the recent results that these (macroscopic) wormhole solutions are unstable \cite{Hertog:2018kbz} in the Euclidean sense. Then the wormholes, despite being smooth, and having a finite action, do not contribute to the path integral.

Many of the non-supersymmetric solutions constructed in \cite{Ruggeri:2017grz} have not been interpreted in this paper. The fate of these solutions is unclear to us. But it could  be that allowing the scalar fields in the dual gauge theories to be turned on, more non-SUSY saddle points exist whose duals are to be found in the supergravity. We leave this for future research.

Our work suggests some extensions. For instance the  Klebanov-Witten background $\rm AdS_5\times T^{1,1}$ has as part of its moduli space the moduli space of $\rm AdS_5\times S^5/\mathbb{Z}_2$ and so the instanton solutions of $\rm AdS_5\times T^{1,1}$ are therefore implicitly contained in this paper. We refer to \cite{Imaanpur:2016beu} for an earlier investigation of the supersymmetric  instantons in $\rm AdS_5\times T^{1,1}$. A less straightforward extension would be to apply these ideas in $\rm AdS_3/CFT_2$ and in particular to the ``D1-D5 CFTs'' dual to $\rm AdS_3 \times S^3\times T^4$ or $\rm AdS_3 \times S^3\times K_3$. The supergravity moduli spaces have not yet been computed but the dual field theory conformal manifolds have been found in \cite{Cecotti:1990kz}. That should be sufficient information to construct the geodesic curves. We hope to come back to this in the future.

\section*{Acknowledgements}
We thank  Nikolay Bobev for useful discussions. The work of SK is supported by the KU Leuven C1 grant ZKD1118 C16/16/005 and by the Belgian Federal Science Policy Office through the Inter-University Attraction Pole P7/37. The work of  TVR is supported by the FWO odysseus grant G.0.E52.14N and by the C16/16/005 grant of the KULeuven. We furthermore acknowledge support from the European Science Foundation Holograv Network  and the COST Action MP1210 `The String Theory Universe'.
\appendix
\section{Parametrizations of complex hyperbolic space \texorpdfstring{$\mathbb{C}H^k$}{CHk}}\label{CHk}
Below we discuss the relevant parametrizations of the complex hyperbolic space $\mathbb{C}H^k$ and define the one which should be most appropriate to the string description.
\paragraph{Fubini - Study coordinates.} These are coordinates $z_\alpha$, $\alpha=0,\dots, k-1 $ in terms of which the K\"ahler potential reads:
\begin{equation}
\mathcal{K}=-2\,\log\left(1-\sum_\alpha |z_\alpha|^2\right)\,.
\end{equation}
This is the parametrization in which the holomorphic prepotential in the projective coordinates $X^\Lambda=(X^{\hat{0}},\, X^\alpha)$ has the form:
\begin{equation}
F(X)=-\frac{i}{2}\,\eta_{\Lambda\Sigma} X^\Lambda X^\Sigma=-\frac{i}{2}\,\left[ (X^{\hat{0}})^2-\sum_\alpha (X^\alpha)^2\right]\,,
\end{equation}
the coordinates being defined as $z^\alpha=X^\alpha/X^{\hat{0}}$, $X^{\hat{0}}\neq 0$. Note that in this parametrization the ${\rm U}(k)$ symmetry is manifest.
\paragraph{Coordinates $t_0,\,v_i$.} Let us split the index $\alpha$ into $0$ and $i=1,\dots, k-1$, and make the following (holomorphic) change of coordinates:
\begin{equation}
z^0=\frac{1+i\,t_0}{1-i\,t_0}\,\,,\,\,\,z_i=-\,\frac{i\, v_i}{1-i\,t_0}\,.
\end{equation}
One can verify that, modulo a K\"ahler transformation, the K\"ahler potential becomes:
\begin{equation}
\mathcal{K}=-2\,\log\left[-2i(t_0-\bar{t}_0)- \sum_i |v_i|^2\right]\,.
\end{equation}
\paragraph{Coordinates $u,\,v_i$.} We can make a further change of coordinates and define:
\begin{equation}
u=t_0-\frac{i}{4}\sum_i v_i^2\,.
\end{equation}
The K\"ahler potential acquires the following form:
\begin{equation}
\mathcal{K}=-2\,\log\left[-2i(u-\bar{u})+ \frac{1}{2} \sum_i (v_i-\bar{v}_i)^2\right]\,.
\end{equation}
This is the parametrization best suited to string theory. It indeed corresponds to the parametrization used in \cite{Louis:2016msm}.
\paragraph{Relation to the solvable coordinates.}
The relation of the above complex coordinates to the solvable ones is the following:
\begin{align}
t_0=-a+i\,\left[e^{2U}+\frac{1}{4}\sum_i\left(\zeta_i^2+\tilde{\zeta}_i^2\right)\right]\,\,,\,\,\,\,
v_i=\tilde{\zeta}_i+i\,{\zeta}_i\,,
\end{align}
or
\begin{align}
u=-\tilde{a}+i\,\left(e^{2U}+\frac{1}{2}\,\sum_i\zeta_i^2\right)=-\tilde{a}+i\, \delta\,,
\end{align}
where, as usual, $\tilde{a}=a-\frac{1}{2}\sum_i \zeta_i\tilde{\zeta}_i$.
Notice that $u$ is precisely the complex modulus we find occurring in the kinetic term and theta term. This is the complex parametrization in which the translational symmetries (real part of $u$ and imaginary part of $v_i$) are manifest.
%

\section{Comparison with the dimensional reduction of Type IIB \texorpdfstring{$T^{1,1}$}{T11}}\label{App:IIB}
In this Appendix we perform an analysis which we believe to be instructive, although not essential to the objectives of the present work.
It is known that direct Kaluza-Klein compactification of the Type IIB theory on a singular background, as the one under consideration, would not give reliable results. We can however address the following question: Had we compactified the ten-dimensional theory on some smoothed version of $AdS_5\times S^5/\mathbb{Z}_k$ with the same topology, what geometry of the moduli space would we have found and how would it have differed from that in (\ref{coset2})?\par  For the sake of concreteness and simplicity, we shall focus on the case $k=2$ and consider, as internal manifold, $T^{1,1}$ as a smoothed version of $S^5/\mathbb{Z}_2$ obtained from a  blow-up of the singular $S^1$ inside the latter \cite{Klebanov:1998hh}. Restricting to those fields which admit an effective five-dimensional supergravity description, the moduli related to the two backgrounds $AdS_5\times S^5/\mathbb{Z}_2$ and $AdS_5\times T^{1,1}$ are the same, since the two manifolds share the same tolopogy. In the latter case the topology fixes the full geometry of the moduli space in the effective five-dimensional description. This however is different from that of (\ref{coset2}). Below we illustrate this difference, along the lines of \cite{Louis:2016msm}, ascribing it to the singular nature of the orbifold.\par
The effective five-dimensional description of the Type IIB theory on $AdS_5\times T^{1,1}$ was considered for instance in \cite{Bena:2010pr,Imaanpur:2016beu,Louis:2016msm}.
In particular the $\sigma$-model describing the $D=10$ axion-dilaton fields and the moduli originating from the RR and NS-NS 2-forms on the 2-cycle of $T^{1,1}$ was worked out within a $D=5$ gauged supergravity. We shall compare our $\sigma$-model metric to that found in \cite{Imaanpur:2016beu,Louis:2016msm} to leading order in the fields $v_I=\tilde{\zeta}_I+ i\,\zeta_I$, and find agreement with our ten-dimensional interpretation of the solvable coordinates. The two metrics differ however by interaction terms which are of higher order in the $v_I$ fields. We shall briefly comment on this, at the end of the present section, following the discussion in \cite{Louis:2016msm}.\par
The type IIB action reads :
\begin{align}
S=\frac{1}{2 \kappa_{10}^2} \int{d^{10}} x e \Bigl( R -\frac{1}{2}\partial_{\hat{\mu}}\phi \partial^{\hat{\mu}}\phi - \frac{e^{2\phi}}{2}\partial_{\hat{\mu}}\chi \partial^{\hat{\mu}}\chi - \frac{e^{-\phi}}{2\cdot 3!}H_{\hat{\mu} \hat{\nu} \hat{\rho}} H^{\hat{\mu}\hat{\nu}\hat{\rho}}  - \frac{e^{\phi}}{2\cdot 3!} \tilde{F}_{\hat{\mu} \hat{\nu} \hat{\rho}} \tilde{F}^{\hat{\mu}\hat{\nu}\hat{\rho}} \nonumber \\
 - \frac{1}{4\cdot 5!} \tilde{F}_{\hat{\mu} \hat{\nu} \hat{\rho} \hat{\sigma} \hat{\delta}} \tilde{F}^{\hat{\mu} \hat{\nu} \hat{\rho} \hat{\sigma} \hat{\delta}} - \frac{1}{2e}\epsilon^{\hat{\mu}_1 \dots \hat{\mu}_{10}} C_{\hat{\mu}_1 \dots \hat{\mu}_4} H_{\hat{\mu}_5 \hat{\mu}_6\hat{\mu}_7} F_{\hat{\mu}_8 \hat{\mu}_9 \hat{\mu}_{10}} \Bigr)\,\,\,,
\end{align}
where $e=\sqrt{|g_{10}|}$,
\begin{equation}
H_{\left(3\right)}\,=\, \frac{H_{\hat{\mu} \hat{\nu} \hat{\rho}}}{3!}\,d x^{\hat{\mu} \hat{\nu} \hat{\rho}}\,\,\,\,\,,\,\,\,\,\,\hat{\mu},\hat{\nu},\hat{\rho}=0,\dots,9
\end{equation}
and
\begin{equation}
H_{\left(3\right)}\,=\,d B_{\left(2\right)}\,\,\,,\,\,\,F_{\left(3\right)}\,=\,d C_{\left(2\right)}\,\,\,,\,\,\,\tilde{F}_{\left(3\right)}\,=\,F_{\left(3\right)}-\chi\,H_{\left(3\right)}\,\,\,.
\end{equation}
Let $\omega$ be the harmonic 2-form corresponding to the 2-cycle of $T^{1,1}$, normalized so that $\omega_{\upalpha \upbeta} \omega^{\upalpha \upbeta}/2=1$, where $\upalpha,\, \upbeta,=1,\dots,5$, only in this appendix, label the $T^{1,1}$ directions.
Let us expand the $2$-forms $B_{\left(2\right)}$ and $C_{\left(2\right)}$ in the basis $(\omega^I)=(\omega)$ of $H^2(T^{1,1},\,\mathbb{Z})$,\footnote{The index $I=1,\dots, k-1$, being $k=2$, has one single value and, in writing $\omega$, we have omitted it.} \emph{i.e.}
\begin{equation}
B_{\left(2\right)}\,=\, b_I \omega^{I}=b\,\omega\,\,\,\,,\,\,\,\,C_{\left(2\right)}\,=\, c_I \omega^{I}=c\,\omega\,.
\end{equation}
Restricting to extremal instantons which do not alter the space-time geometry, we end up with the effective five-dimensional action (recall that $k=2$):
\begin{align}
S=\frac{1}{2 k_5^2} \int_{{\rm AdS}_5}{d^{5}} x\, \sqrt{|g_5|}\, \Bigl[ R -\frac{1}{2}\partial_{\mu}\phi \partial^{\mu}\phi - \frac{e^{2\phi}}{2}\partial_{\mu}\chi \partial^{\mu}\chi - \frac{e^{-\phi}}{2} \partial_{\mu} b_I \partial^{\mu} b_I  - \frac{e^{\phi}}{2}\left(\,\partial_{\mu} c_I-\chi\,\partial_{\mu} b_I\,\right)^2+\dots \Bigr]\,\,\,\,\,.
\end{align}
The $\sigma$-model metric for the moduli $\{\,\phi,\,\chi\,,b_I\,,c_I\,\}$ has the following general structure
\begin{equation}
ds^2\,=\, \left(d\phi\right)^2\,+\,e^{2\phi}\,\left(d\chi\right)^2\,+\,{e^{-\phi}}\,\sum_{I=1}^{k-1}\left(d b_I\right)^2\,+\,{e^{\phi}}\,\sum_{I=1}^{k-1}\left(d c_I\,-\,\chi\,d b_I\,\right)^2\,\,\,.
\end{equation}
Note the $2(k-1)$ translational isometries of the above metric:
\begin{equation}
b_I\rightarrow b_I+ \xi_I\,\,,\,\,\,\,b_I\rightarrow b_I+ \lambda_I\,\,,\,\,\,\,\xi_I,\,\lambda_I={\rm const.}
\end{equation}
only half of which are present in our $\sigma$-model metric (\ref{metric}).\par
The $(2k=4)$-dimensional moduli space so obtained is a K\"ahler manifold with complex coordinates
\begin{equation}
z_1\,=\,\chi\,+\,i e^{-\phi}\quad,\quad z_{I+1}\,=\,c_I\,-\,z_1 b_I\quad,\quad I=1,\dots,k-1
\end{equation}
and K\"ahler potential
\begin{equation}
\mathcal{K}\,=\,-2\,\ln\left(\,z_1 - \bar{z}_1\,\right)\,-\frac{i}{2}\,\sum_I\,\frac{\left(\,z_{I+1} - \bar{z}_{I+1}\,\right)^2}{z_1 - \bar{z}_1}\label{K1}\quad.
\end{equation}
This is the moduli-space geometry found in \cite{Louis:2016msm} from an analysis of the corresponding $D=5$ gauged supergravity, and in \cite{Imaanpur:2016beu} from dimensional reduction.
Our moduli space $SU\left(1,k\right)/U\left(k\right)$ is instead associated with a K\"ahler potential
\begin{equation}
\mathcal{K}\,=\,-2\,\ln\left[-2i\,\left(\,u - \bar{u}\,\right)\,+\frac{1}{2}\sum_I\,\left(\,v_{I} - \bar{v}_{I}\right)^2\,\right]\quad.
\end{equation}
Let us rescale $v_I \rightarrow\sqrt{\epsilon}\,v_I$ and keep only terms of $\mathcal{O}\left(\epsilon\right)$
\begin{equation}
\mathcal{K}\,=\,-2\,\ln\left[-2i\,\left(\,u - \bar{u}\,\right)\,\right]\,-\,\frac{i\,\epsilon}{2}\,\sum_I\,\frac{\left(\,v_{I} - \bar{v}_{I}\right)^2}{\left(\,u - \bar{u}\,\right)}\label{K2}\quad.
\end{equation}
Reabsorbing $\sqrt{\epsilon}$ in $v_I$ (which means considering $|v_I|$ ``small''), and
identifying (\ref{K1}) with (\ref{K2}) we get
\begin{align}
u\,&=\,z_1\,=\,\chi\,+\,i\,e^{-\phi}\,=\,-\tilde{a}\,+\,i\,\delta\quad,\\
v_I \,&=\,\tilde{\zeta}_I \,+i\,{\zeta}_I \,=\,-\left(\,c_I \,-\,u\,b_I \,
\right)\,=\,-\left(\,c_I \,-\,\chi\,b_I \,\right)\,+i\,e^{-\phi}\,b_I \quad,
\end{align}
\begin{equation}
\begin{cases}
\zeta_I \,=\,e^{-\phi}\,b_I \\
\tilde{\zeta}_I \,=\,-\,\left(\,c_I \,-\,\chi\,b_I \,\right)\\
\delta\,=\,e^{2U}\,+\,\frac{1}{2}\,\sum_I \,\left(\,\zeta_I \,\right)^2\,=\,e^{-\phi}\\
\tilde{a}\,=\,-\chi\quad.
\end{cases}
\end{equation}
The expansion in $\epsilon$, necessary to connect the two moduli spaces, was interpreted in \cite{Louis:2016msm} as deriving from an expansion in the Newton's constant $\kappa_5$ which naturally multiplies the $b,\,c$ scalars and which, in the large-N limit, is small. The singular orbifold $S^5/\mathbb{Z}_2$ is however obtained by sending to zero the blow-up parameter. We think it is reasonable to assume that the presence, in the effective supergravity description, of these two competing limits justifies the existence in the moduli space-geometry of the orbifold reduction, as described by the effective $D=5$ gauged supergravity, of interaction terms which are absent in the $T^{1,1}$ case. We do not expand on this issue further in this paper.

\section{On the solvable parametrization of \texorpdfstring{$\mathbb{C}H^k$}{CHk} and its manifest symmetry}\label{Solvpar}
Referring to the Appendix B of \cite{Ruggeri:2017grz}, we write
\begin{align}
\mathfrak{su}(1,k)\,=\,\left[\,\mathfrak{sl}(2)\,\oplus\,\mathfrak{u}(k-1)\,\right]\,\oplus_S\,\left[\,({\bf 2},{\bf k-1})_{{\bf +1}}\,+\,({\bf 2},{\bf k-1})_{{\bf -1}}\,\right]\,,
\end{align}
where the grading refers to the ${\rm U}(1)$ inside ${\rm U}(k-1)$ in the Wick-rotated group of
\begin{align}
\mathfrak{sl}(1+k)\,=\,\left[\,\mathfrak{sl}(2)\,\oplus\,\mathfrak{gl}(k-1)\,\right]\,\oplus_S\,\left[\,({\bf 2},{\bf k-1})_{{\bf +1}}\,+\,({\bf 2},{\bf k-1})_{{\bf -1}}\,\right]\,,
\end{align}
being $\mathfrak{sl}(2)$ the isometry algebra of ${\rm SL}(2)/{\rm SO}(2)$ parametrized by $U$ and $a$ .\\
To define the solvable Lie algebra we further split $\mathfrak{sl}(2)$
\begin{align}
\mathfrak{sl}(2)\,\,\,\rightarrow\,\,\,{\bf 1_0}\,+\,{\bf 1_{+2}}\,+\,{\bf 1_{-2}}
\end{align}
and use a double grading structure with respect to $H_0\,\in\,\mathfrak{sl}(2)$ parametrized by $U$ and $\mathfrak{gl}(1)$ in $\mathfrak{gl}(k-1)$ . We shall consider, for the sake of simplicity, the Wick-rotated case, obtaining
\begin{align}
\mathfrak{sl}(1+k)\,&=\,\mathfrak{gl}(k-1)_{({\bf 0},{\bf 0})}\,\oplus\,{\bf 1}_{({\bf 0},{\bf 0})}\,\oplus\,\bigl[\,{\bf 1}_{({\bf 2},{\bf 0})}\,\oplus\,{\bf 1}_{({\bf -2},{\bf 0})}\,\oplus\,({\bf k-1})_{({\bf +1},{\bf +1})}\,\nonumber\\
&\oplus\,({\bf k-1})_{({\bf +1},{\bf -1})}\,\oplus\,({\bf k-1})_{({\bf -1},{\bf +1})}\,\oplus\,({\bf k-1})_{({\bf -1},{\bf -1})}\bigr]\,,
\end{align}
where the solvable Lie algebra, $Solv$, is generated by
\begin{align}
H_0\,=\,{\bf 1}_{({\bf 0},{\bf 0})}\quad,\quad T_\bullet\,=\,{\bf 1}_{({\bf 2},{\bf 0})}\quad,\quad &T_i\,+\,T^i\,=\,({\bf k-1})_{({\bf +1},{\bf +1})}\quad,\nonumber\\
&T_i\,-\,T^i\,=\,({\bf k-1})_{({\bf +1},{\bf -1})}\,.
\end{align}
We see that the manifest symmetry is generated by $\mathfrak{gl}(k-1)$ which acts linearly on $\zeta^i$, $\tilde{\zeta}_i$ and has no action on $U$ and $a$ . \\
The linear action of a generic element of $\mathfrak{gl}(k-1)$ on the scalars $\zeta^i$,$\tilde{\zeta}_i$ is described by the following matrix:
\begin{align}
\mathbbm{J}\,=\,\left(
                  \begin{array}{c | c}
                    {{\bf J}^{i}}_{j} &  {\bf K}^{i j} \\ \hline
                    {\bf K}_{i j} & {{\bf J}_i}^{j}  \\
                  \end{array}
                \right)\quad,\quad \begin{array}{c }  {\bf K}^{i j}\,=\, {\bf K}_{i j}\,=\, {\bf K}^{j i}\\
                {{\bf J}^{i}}_{j}\,=\,-{{\bf J}_{j}}^{i} \,=\,{{\bf J}_{i}}^{j}\end{array}\quad,
\end{align}
where $J$ generates $SO(k-1)\subset GL(k-1)$. The matrix $\mathbbm{J}$ defines the embedding of $GL(k-1)$ within $ Sp(2(k-1))$.\par
Defining ${\bf J}\equiv ({{\bf J}^{i}}_{j})$ and $ {\bf K}\equiv  {\bf K}_{i j}$, we can write $\mathbbm{J}$ in the form of a Kronecker product
\begin{align}
\mathbbm{J}\,=\,{\bf J}\,\otimes\,\mathbbm{1}_2\,&+\, {\bf K}\,\otimes\,\sigma_1\quad,\\
\nonumber\\
\mathbb{C}\,=\,\mathbbm{1}\,\oplus\,i\sigma_2\,=\,\left(
                                                     \begin{array}{c | c}
                                                       0 & \mathbbm{1} \\ \hline
                                                       -\mathbbm{1} & 0 \\
                                                     \end{array}
                                                   \right)
\quad&,\quad \eta\,=\,\mathbbm{1}\,\oplus\,\sigma_3\,=\,\left(
                                                     \begin{array}{c | c}
                                                       \mathbbm{1} & 0 \\ \hline
                                                       0 & -\mathbbm{1} \\
                                                     \end{array}
                                                   \right)\quad;
\end{align}
we observed that
\begin{align}
\mathbbm{J}^{T}\,\mathbb{C}\,&=\,-\mathbb{C}\,\mathbbm{J}\,\,\,,\\
\mathbbm{J}^{T}\,\eta\,&=\,-\eta\,\mathbbm{J}\,\,\,.
\end{align}
Using the above properties, one can easily verify that the terms $\mathcal{Z}^M \mathbb{C}_{MN} d\mathcal{Z}^N$ and $d\mathcal{Z}^M \eta_{MN} d\mathcal{Z}^N$ in the metric are indeed manifestly invariant under the action of $GL(k-1)$ through the matrix $\mathbbm{J}$.\\
When we make contact with the IIB string theory, however, we have the following correspondences
\begin{equation}
e^{-\phi}\,=\,e^{2U}\,+\,\frac{1}{2}\,\sum_i\,\left(\,\zeta^i\,\right)^2\quad,\quad\tilde{a}\,=\,a\,-\,\frac{1}{2}\,\sum_i\,\tilde{\zeta}_i\,\zeta^i\,=\,-\chi\,,
\end{equation}
which are invariant only under $O(k-1)$, while the non-compact transformations in $GL(k-1)$ act non linearly on $e^{-\phi}$ and $\chi$ . It follows that the parametrization of the moduli space which is directly related to the string excitations has manifest $O(k-1)$ symmetry.
We end this Appendix by giving the general expression of the Noether currents associated with the solvable generators $T_{\mathcal{I}}$:
\begin{equation}
J_{\mathcal{I}|\mu}(\phi)\,\equiv \, \frac{1}{2\kappa_5^2}\,K_{\mathcal{I}}{}^{\mathcal{J}}(\phi)\,G_{\mathcal{JK}}(\phi)\,\partial_\mu \phi^{\mathcal{K}}\,,\label{NC}
\end{equation}
where
\begin{equation}
K_{\mathcal{I}}(\phi)\equiv K_{\mathcal{I}}{}^{\mathcal{J}}(\phi)\,\frac{\partial}{\partial \phi^{\mathcal{J}}}\,,\label{Kill}
\end{equation}
are the Killing vectors corresponding to $T_{\mathcal{I}}$.
Note that the Killing vectors associated with the axion fields $\chi^{\underline{\alpha}}$ defined in (\ref{axioncollective}) are simply:
\begin{equation}
K_{\underline{\alpha}}=\frac{\partial}{\partial \chi^{\underline{\alpha}}}\,.
\end{equation}

\bibliographystyle{utphys}
{\small
\bibliography{refs}}

\end{document}